%% file: main_STTT.tex
\RequirePackage{fix-cm}
\documentclass[twocolumn]{svjour3}          

\input{ancillary/packages}

\smartqed 
\usepackage{graphicx}

\input{ancillary/macros}

\begin{document}

\title{Scenario-Based Verification of Uncertain Parametric  MDPs
\thanks{This work was partially funded by NWO grant NWA.1160.18.238 (PrimaVera), the ERC AdG 787914 (FRAPPANT), NSF 1652113, ONR N000141613165, NASA NNX17AD04G and AFRL FA8650-15-C-2546.}
}

\author{Thom Badings \and Murat Cubuktepe \and Nils Jansen \and Sebastian Junges \and Joost-Pieter Katoen \and Ufuk Topcu
}

\institute{T.\ Badings, N.\ Jansen and S.\ Junges \at
              Radboud University, Nijmegen, the Netherlands \\
              \email{thom.badings@ru.nl}
           \and
           M. Cubuktepe and U.\ Topcu \at
              The University of Texas at Austin, Austin, TX, USA
           \and
           J.P.\ Katoen \at
              RWTH Aachen University, Aachen, Germany
}

\date{Received: June 17, 2022}

\maketitle

\begin{abstract}
We consider parametric Markov decision processes (pMDPs) that are augmented with unknown probability distributions over parameter values.
The problem is to compute the probability to satisfy a temporal logic specification with any concrete MDP that corresponds to a sample from these distributions.
As solving this problem precisely is infeasible, we resort to sampling techniques that exploit the so-called scenario approach.
Based on a finite number of samples of the parameters, the proposed method yields high-confidence bounds on the probability of satisfying the specification. 
The number of samples required to obtain a high confidence on these bounds is independent of the number of states and the number of random parameters. 
Experiments on a large set of benchmarks show that several thousand samples suffice to obtain tight and high-confidence lower and upper bounds on the satisfaction probability.
\end{abstract}


\input{sections/1_introduction}
\input{sections/2_motivating}
\input{sections/3_preliminaries}
\input{sections/4_problem}
\input{sections/5_robust}
\input{sections/6_scenario_reach}
\input{sections/7_experiments}
\input{sections/8_related}
\input{sections/9_conclusion}

\bibliographystyle{spmpsci}
\bibliography{literature}

\appendix
\input{sections/A_proofs}

\end{document}

%% file: ancillary/packages.tex
\usepackage{amsmath,amssymb,amsfonts}
\usepackage{mathtools}

\usepackage{amsthm}
\usepackage[dvipsnames]{xcolor}
\usepackage{xspace}
\usepackage{microtype}
\usepackage{colonequals}

 \usepackage{amssymb, bm}

\usepackage{multirow}

\usepackage{paralist}
\usepackage{caption}
\usepackage{subcaption}
\usepackage{url}
\usepackage{appendix}
\usepackage{nicefrac}
\usepackage{tikz} 
\usepackage{mdframed}

\usepackage{hyperref}

\usepackage{tikz}
\usepackage{pgfplots,pgfplotstable}
\usetikzlibrary{arrows,decorations.pathmorphing,positioning,fit,trees,shapes,shadows,automata,calc} 
\usetikzlibrary{patterns,arrows,arrows.meta,calc,shapes,shadows,decorations.pathmorphing,decorations.pathreplacing,automata,shapes.multipart,positioning,shapes.geometric,fit,circuits,trees,shapes.gates.logic.US,fit,automata,decorations,shapes.geometric}
\usepgfplotslibrary{fillbetween}
\pgfplotsset{compat=1.8}

\usepackage{program}%

%% file: ancillary/macros.tex
\newcommand{\eg}{e.g.\@\xspace}

\newcommand{\tool}[1]{\textrm{#1}\xspace}

\newcommand{\subjectto}{\textnormal{subject to}}

\DeclareMathOperator*{\maximize}{maximize}
\DeclareMathOperator*{\minimize}{minimize}
\usepackage{array,booktabs}
\newcolumntype{C}{>{$\displaystyle}c<{$}}
\usepackage[normalem]{ulem}
\useunder{\uline}{\ul}{}

\newcommand{\PP}{\mathbb{P}}

\newcommand{\RR}{\mathbb{R}}

\newtheorem{coro}{Corollary}

\newtheorem{assume}{Assumption}

\newcommand{\dtmc}{\mathcal{D}}

\renewcommand{\p}{\ensuremath{\mathbf{P}}}
\newcommand{\probdist}{\ensuremath{\mathbb{P}}}

\newcommand{\violating}{\ensuremath{N_{\neg\varphi}}}
\newcommand{\satisfying}{\ensuremath{N_{\varphi}}}

\newcommand{\reachProp}[2]{\ensuremath{\p_{\geq #1}(\finally #2)}}
\newcommand{\reachPropg}[2]{\ensuremath{\p_{\leq #1}(\finally #2)}}

\newcommand{\satprob}{F(\vmdp,\varphi)}
\newcommand{\satreachprob}{F(\vmdp,\varphi)}

\newcommand{\falseprob}{F(\vmdp,\neg\varphi)}

\newcommand{\reachPropgT}{\ensuremath{\reachPropg{\lambda}{T}}}
\newcommand{\reachPropgTvary}{\ensuremath{\reachPropg{\lambda^*(\mathcal{U}_N)}{T}}}

\newcommand{\finally}{\lozenge}

\definecolor{red}{rgb}{0.745,0.192,0.102}

\renewcommand{\R}{\mathbb{R}}

\newcommand{\Paramvar}{\ensuremath{{V}}\xspace}        

\newcommand{\sinit}{s_{\mathit{I}}} 
\newcommand{\mdp}{\mathcal{M}}
\newcommand{\pmdp}{\mdp}
\newcommand{\vmdp}{\mdp_{\probdist}}

\newcommand{\probmdp}{\mathcal{P}}

\newcommand{\paramspace}[1][]{\ensuremath{\mathcal{V}_{#1}}}

\renewcommand{\Pr}{\ensuremath{\textnormal{Pr}}}
\newcommand{\prob}{\mathsf{\mathbf{Pr}}}

\newcommand{\sched}{\ensuremath{\sigma}}
\newcommand{\Sched}{\ensuremath{\mathit{Str}}}
\newcommand{\sol}{\ensuremath{\mathsf{sol}}}

\newcommand{\Act}{\ensuremath{\mathit{Act}}}
\newcommand{\ActS}{\ensuremath{\mathit{ActS}}}

\newcommand{\act}{\ensuremath{\alpha}}

\DeclareMathAlphabet{\mathpzc}{OT1}{pzc}{m}{it}
\def\presuper#1#2%
  {\mathop{}%
   \mathopen{\vphantom{#2}}^{#1}%
   \kern-\scriptspace%
   #2}

\newcommand{\confidence}{\ensuremath{\beta}}
\newcommand{\violation}{\ensuremath{\epsilon}}

\newcommand{\cardQ}{\ensuremath{\vert \mathcal{Q} \vert}}

%% file: sections/1_introduction.tex
\section{Introduction}
\label{sec:introduction}

\paragraph{MDPs.} Markov decision processes (MDPs) model sequential decision-making problems in stochastic dynamic environments~\cite{puterman2014markov}. 
They are widely used in areas such as planning~\cite{russell2016artificial}, reinforcement learning~\cite{sutton2018reinforcement}, formal verification~\cite{BK08}, and robotics~\cite{mcallister2012motion}.
Mature model checking tools such as \tool{PRISM}~\cite{KNP11} and \tool{Storm}~\cite{DBLP:conf/cav/DehnertJK017} employ efficient algorithms to verify the correctness of MDPs against temporal logic specifications~\cite{Pnu77} provided all transition probabilities and cost functions are exactly known. 
In many applications, however, this assumption may be unrealistic, as certain system parameters are typically not exactly known and under control by external sources.

\paragraph{Uncertainty on MDPs.}
A common approach to deal with unknown system parameters is to describe transition probabilities of an MDP using intervals~\cite{DBLP:conf/lata/DelahayeLLPW11,puggelli2013polynomial,givan2000bounded} or generalizations to a class of uncertain MDPs~\cite{DBLP:conf/cdc/WolffTM12,nilim2005robust,wiesemann2013robust}. 
Solution approaches rely on the limiting assumption that the uncertainty at different states of the MDP is independent from each other. 
As an example, consider a simple motion planning scenario where an unmanned aerial vehicle (UAV) is tasked to transport a certain payload to a target location. 
The problem is to compute a \emph{strategy} (or \emph{policy}) for the UAV to successfully deliver the payload while taking into account the weather conditions. 
External factors such as the wind strength or direction may affect the movement of the UAV.
The assumption that such weather conditions are independent between the different possible states of UAV is unrealistic and may yield  pessimistic verification results. 

\paragraph{Illustrative examples.}
We stress that the same situation appears in various systems.
For example, in the verification of network protocols, we typically do not precisely know the channel quality (i.e., the loss rate). 
However, the loss rate is independent of the question of whether we are, e.g., probing or actually sending useful data over the network. 
A typical verification task would be to show that the protocol yields a sufficiently high quality of service.  
A verification approach that pessimistically assumes that the channel quality depends on the protocol state may be too pessimistic and fail to establish that the protocol provides the required quality of service.

\paragraph{Parametric models.}
Parametric Markov models allow to explicitly describe that some probabilities are unknown but explicitly related~\cite{Daw04,param_sttt,DBLP:journals/corr/abs-1903-07993}.
In a parametric MDP, one uses variables (\emph{parameters}) that encode, e.g., the probability of wind gusts affecting a UAV, or the probability of packet loss in a network.
Transition probabilities are then given as expressions over these parameters.
Substituting the parameters with concrete values yields an \emph{induced} (standard) MDP. 
A variety of parameter synthesis methods have been devised, see the related work in \autoref{sec:related} for details. 
A typical verification query concerns feasibility, that is, \emph{whether there exist parameter values such that the induced model satisfies a specification}, which implicitly assumes that the parameters are controllable. 
Another query is to ask \emph{whether for all parameter values the induced model satisfies a specification}. 
The latter can lead to pessimistic verification results: a UAV may be able to fly during most weather conditions, but it may be impossible to find a satisfying strategy for flying during a rare storm.

\paragraph{Uncertain parametric models.}
Rather than asking about the \emph{existence} of parameter values, we want to analyze a system by considering the \emph{typical} parameter values. 
In terms of our examples, this means that we want to investigate the typical weather conditions and the typical channel qualities. 
Similar to \cite{DBLP:conf/valuetools/Scheftelowitsch17}, we, therefore, assume that the parameters are random variables.
For instance, weather data in the form of probability distributions may provide additional information on potential changes during the mission, or a measurement series may provide typical channel qualities.
For weather data, such probability distributions may be derived from historical data of, for example, the wind speed~\cite{Papaefthymiou2008}.

\paragraph{Problem statement.}
We study a setting where the uncertain parameters are random variables that are defined on an arbitrary (joint) probability space over all parameters. 
We assume that we can sample \emph{independent and identically distributed} parameter values from this distribution and solve the following problem.

\begin{mdframed}[backgroundcolor=gray!30, nobreak=true]
    \textbf{Problem statement.} \, Given a parametric MDP and a distribution over the parameter values, compute the probability with which  any randomly drawn parameter values yield an induced MDP that satisfies a given specification.
\end{mdframed}

We call this probability the \emph{satisfaction probability}.
The intuition is that the question of whether all (or some) parameter values satisfy a specification -- as is often done in parameter synthesis~\cite{DBLP:journals/corr/abs-1903-07993} -- is replaced by the question of \emph{how much we expect the (sampled) model to satisfy a specification}.
For example, a satisfaction probability of $80\%$ tells that, if we randomly sample the parameters, with a probability of $80\%$ there exists a strategy for the resulting MDP satisfying the specification. Importantly, we thus assume that the parameter values are \emph{observable}, and hence known when synthesizing a strategy. In every concrete MDP, we may use a different strategy. 
This is in contrast to a robust strategy synthesis approach, where a single strategy is sought that is robust against all (or a portion of the) parameter valuations.

\paragraph{Scenario-based verification.}
In this paper, we devise a method that answers the problem statement up to a user-specified confidence level. 
That is, we aim to solve the problem statement up to a statistical guarantee. 
To achieve this, we resort to \emph{sampling-based} algorithms that yield a confidence (probability) on the bounds of the satisfaction probability.
In doing so, \emph{we do not make any assumptions} about the distribution over the parameter values.
Referring back to the UAV example, we want to compute a confidence bound on the probability for the UAV to successfully finish its mission for some strategy.
To derive confidence bounds, we first formulate the problem of (exactly) computing the satisfaction probability as a \emph{chance-constrained optimization program}.
However, this problem is very hard to solve~\cite{campi2018introduction}, especially because we do not assume any knowledge on the probability distribution of the parameters.
We, therefore, use a technique known as \emph{scenario optimization} (also called the \emph{scenario approach}), which provides guarantees on the satisfaction probability via sampling techniques~\cite{DBLP:journals/siamjo/CampiG08,campi2011sampling}.
The basic idea is to consider a finite set of samples from the distribution over the parameters and restrict the problem to these samples only.
This so-called \emph{scenario optimization problem} can be solved efficiently~\cite{boyd_convex_optimization}.
The solution to the scenario program is, \emph{with a high confidence probability}, a solution to the previously mentioned chance-constrained program.

\paragraph{Our approach.}
For our setting, we first sample a finite number of parameter instantiations, each of which induces a concrete MDP.
We can check the satisfaction of the specification for these concrete MDPs efficiently using, \eg, a probabilistic model checker.
Based on the results, we compute an estimate of the satisfaction probability, which is a lower bound on the true satisfaction probability with the desired confidence probability.
For example, we may obtain a lower bound on the satisfaction probability of $80\%$, which holds with a confidence probability of at least $90\%$.
We show that the probability of an incorrect lower bound on the satisfaction probability diminishes to zero \emph{exponentially rapidly} with an increasing sample size.
Moreover, the number of required samples depends on neither the size of the state space nor the number of random parameters.
Finally, we show that we can use the same technique to additionally compute \emph{upper bounds} on the satisfaction probability.

\paragraph{Empirical evaluation.}
In our experiments, we validate the theoretical results using several MDPs 
that have different sizes of state and parameter spaces.
We demonstrate experimentally that the required number of parameter samples is indeed not sensitive to the dimension of the state and parameter space.
In addition, we show the effectiveness of our method with a new dedicated case study based on the aforementioned UAV example which incorporates 900 random parameters. 

\paragraph{Contributions.}
This paper revises an earlier conference paper~\cite{DBLP:conf/tacas/Cubuktepe0JKT20} as follows.
Due to new results in~\cite{GarattiCampi2021} that lift some assumptions required for the scenario approach, we can simplify and generalize our approach by a simplified chance-constrained program. 
This program can be used in conjunction with all (standard temporal) properties on parametric MDPs. 
This change in the approach yields completely revised technical sections of the paper. Furthermore, this paper fixes a technical error in~\cite{DBLP:conf/tacas/Cubuktepe0JKT20}. 
The (new) bounds in Theorem 1 of this paper are less pessimistic. 
The (new) bounds in Theorem 2 are now correct at the cost of being slightly more conservative.

%% file: sections/2_motivating.tex
\section{Motivating Example}
\label{sec:Example}
We consider the previously mentioned UAV motion planning example in more detail, where the objective is to transport a payload from one end of a valley to the other. 
A specification for the UAV would be that it (with at least probability $x$) realizes this objective. 
The typical approach to verify the UAV against this specification is to create a model that faithfully captures its dynamics.

However, the dynamics of the UAV depend on the weather conditions, which may be different on each day.
Thus, each weather condition induces a distinct model for the UAV.
In line with the problem statement, we assume that the weather conditions are deterministically observed on the day itself, and we can adapt the strategy accordingly.
When designing the UAV, we may require that the expected number of days per year on which the UAV can satisfy a mission objective is sufficiently high. 
Concretely, this translates to the requirement that the UAV shall satisfy the specification above on, e.g., at least $90\%$ of the days.
More abstractly, this requirement implies we want to show that the probability of a random day yielding weather conditions on which a specification of the corresponding model is satisfied is at least $90\%$. 
To this end, we assume that we have historical data that describes a distribution over weather conditions. 

\begin{figure}[t]
\centering
\scalebox{0.9}{
\input{tikz/UAV-plan}
}
\caption{An example of a 3D UAV benchmark with obstacles (red boxes) and a target area (green box).}
\label{fig:drone}
\end{figure}
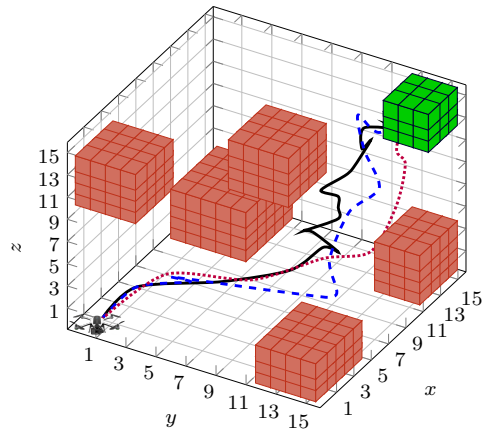

\paragraph{Model construction.}
Planning scenarios like the UAV example are naturally modelled by MDPs, where the actions determine the high-level control actions and the action outcomes are stochastic due to the influences of the environment. 
While this planning problem is, to some degree, continuous, high-level planners often discretize the world. 
We thus obtain the following grid world in which the UAV can decide to fly in either of the six cardinal directions (N, W, S, E, up, down).
States encode the position of the UAV, the current weather condition (sunny, stormy), and the general wind direction in the valley. 
In this particular scenario, we assume that the probabilistic outcomes are (only) determined by the wind in the valley and the control action. 
Concretely, we assume that an action moves the UAV one cell in the corresponding direction.
Moreover, the wind moves the UAV one cell in the wind direction with a probability $p$, which depends on the wind speed. 
Furthermore, we assume that the weather and wind-conditions change during the day and are described by a stochastic process. 

We observe that some probabilities in the system are not fixed but rather a function of the weather. 
Thus, the model is an uncertain MDP (uMDP) whose transition probabilities depend on the weather. 
Concretely, parameters describe how the weather affects the UAV in different zones of the valley, and how the weather/wind may change during the day.
Historical weather data now induces a distribution over the (joint) parameters. 
Sampling from this distribution yields a concrete instantiated MDP. 
The problem is to compute the satisfaction probability, i.e., the probability that \emph{for any sampled MDP, we are able to synthesize a UAV strategy that satisfies the specification.}
Fig.~\ref{fig:drone} shows an example environment for the UAV, with the target zone in green and zones to avoid shown in red.
The shown trajectories are typical paths under three different weather conditions (we refer to the experiments in \autoref{sec:numerical_examples} for details).

%% file: tikz/UAV-plan.tex
\begin{tikzpicture}
\begin{axis}[
	view={120}{40},
width=2.917in,
height=2.95in,
	grid=major,
	xmin=-0.9,xmax=14.4,
	ymin=-0.9,ymax=14.4,
	zmin=-0.9,zmax=14.4,
	enlargelimits=upper,
	xtick={1,3,...,15},
	ytick={1,3,...,15},
	ztick={1,3,...,15},
	xlabel={$x$},
	ylabel={$y$},
	zlabel={$z$},
	x dir=reverse,
	point meta={x},
	colormap={summap}{
		color=(red); color=(red); 
		color=(red); color=(red) 
		color=(red) color=(red) 
		color=(red)
	},
	scatter/use mapped color={
		draw=mapped color,fill=mapped color!70},
	]


	\addplot3[only marks,scatter,draw=none,fill=red!80!black,mark=cube*,mark size=6.6,opacity=1] 		table {results/UAV-STTT/obstacles.dat};
	
	\addplot3[only marks,draw=blue!20!black,fill=green!80!black,mark=cube*,mark size=6.6] 		table {results/UAV-STTT/goal.dat};
	
	\addplot3[draw=black,smooth,very thick,clip marker paths=true] table[x=x,y=y,z=z,opacity=10] {results/UAV-STTT/data_uniform.dat};
				
	\addplot3[draw=blue,dashed,smooth,very thick,clip marker paths=true] table[x=x,y=y,z=z,opacity=10] {results/UAV-STTT/data_x_neg.dat};
		
	\addplot3[draw=purple,densely dotted,smooth,very thick,clip marker paths=true] table[x=x,y=y,z=z,opacity=10] {results/UAV-STTT/data_y_pos.dat};
					

	  \node[]  at (930,070) {\includegraphics[height=0.8cm,width=0.8cm]{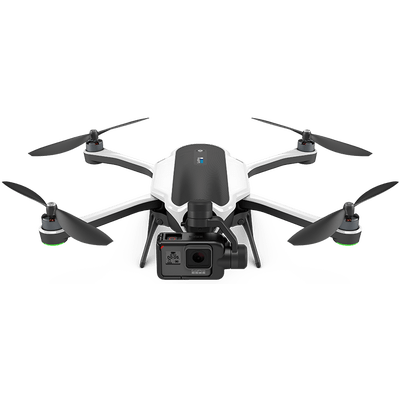}};

\end{axis}
\begin{axis}[
	view={120}{40},
width=2.917in,
height=2.95in,
hide x axis,
hide y axis,
hide z axis,
	grid=major,
	xmin=-0.2,xmax=9.0,
	ymin=-0.2,ymax=8.9,
	zmin=-0.2,zmax=8.9,
	enlargelimits=upper,
	xtick={1,3,...,9},
	ytick={1,3,...,9},
	ztick={1,3,...,9},
	xlabel={$x$},
	ylabel={$y$},
	zlabel={$z$},
	point meta={x},
	colormap={summap}{
		color=(red); color=(red); 
		color=(red); color=(red) 
		color=(red) color=(red) 
		color=(red)
	},
	scatter/use mapped color={
		draw=mapped color,fill=mapped color!70},
	]

	



        

	  \node[]  at (930,070) {\includegraphics[height=0.8cm,width=0.8cm]{figures/uav.png}};

\end{axis}

\end{tikzpicture}

%% file: sections/3_preliminaries.tex
\section{Preliminaries}

In the following, we use probability distributions over finite and infinite sets, for which we refer to~\cite{bertsekas2000introduction} for details.
Let $\Paramvar= \lbrace x_1,\ldots,x_n\rbrace$ be a finite set of variables \emph{(parameters)} over $\mathbb{R}^n$. 
The set of polynomials over $\Paramvar$, with rational coefficients, is denoted by $\mathbb{Q}[\Paramvar].$ 
We denote the cardinality of a set $\mathcal{U}$ by $\vert \mathcal{U} \vert$. 

\subsection{Parametric Models}
We introduce parametric Markov decision processes. 
Note that we omit reward models, but our methods are directly applicable to reward measures.

\begin{definition}[pMDP]
	A \emph{parametric Markov decision process} (pMDP) $\pmdp$  is a tuple $\pmdp = (S, \Act, \sinit, \Paramvar, \mathcal{P})$
	with a finite set $S$ of \emph{states}, a finite set $\Act$ of \emph{actions}, an \emph{initial state} $\sinit \in S$, a finite set $\Paramvar$ of \emph{parameters}, and a \emph{transition function} $\probmdp \colon S \times \Act \times S \rightarrow \mathbb{Q}[\Paramvar]$.
\end{definition}

\noindent
The set $\ActS(s)$ of \emph{enabled} actions at state $s \in S$ is $\ActS(s) = \{\act \in \Act \mid \exists s'\in S,\,\probmdp(s,\,\act,\,s') \neq 0 \}$.
Without loss of generality, we require $\ActS(s) \neq \emptyset$ for all $s\in S$.
If $\vert \ActS(s) \vert = 1$ for all $s \in S$, $\mdp$ is a \emph{parametric discrete-time Markov chain (pMC)} and we denote its transition function by $\probmdp(s,s') \in \mathbb{Q}[\Paramvar]$.

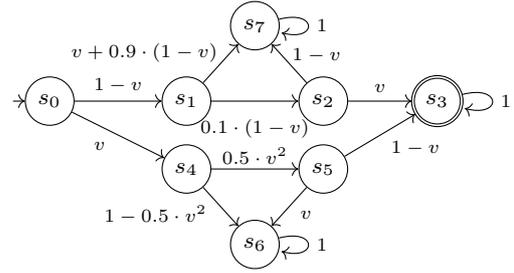
\begin{figure}[t]
\centering
\scalebox{1.0}{

\begin{tikzpicture}[scale=0.4, nodestyle/.style={draw,circle},baseline=(s6)]
    \node [nodestyle,initial,initial text=] (s0) at (0,0) {$s_0$};
    \node [nodestyle] (s1) [on grid, right=1.8cm of s0] {$s_1$};
    \node [nodestyle] (s2) [on grid, right=1.8cm of s1] {$s_2$};
    \node [nodestyle ,accepting] (s3) [on grid, right=1.5cm of s2] {$s_3$};
    \node [nodestyle] (s4) [on grid, below=0.9cm of s1] {$s_4$};
        \node [nodestyle] (s5) [on grid, right=1.8cm of s4] {$s_5$};
    \node [nodestyle] (s6) [on grid,below=1.0cm of s4,xshift=0.9cm] {$s_6$};
        \node [nodestyle] (s7) [on grid,above=1.0cm of s1,xshift=0.9cm] {$s_7$};

    \draw[->] (s0) -- node [auto] {\scriptsize $1-v$} (s1);
    \draw[->,] (s0) -- node [below,pos=0.3, yshift=-0.1cm] {\scriptsize $v$} (s4);

    \draw[->] (s1) -- node [below,yshift=-3.5pt] {\scriptsize $0.1\cdot(1-v)$} (s2);
        \draw[->] (s2) -- node [right,xshift=-2pt,yshift=3pt] {\scriptsize $1-v$} (s7);
        \draw[->] (s1) -- node [left,xshift=2pt,yshift=4pt] {\scriptsize $v+0.9\cdot(1-v$)} (s7);
        \draw[->] (s5) -- node [right,xshift=1pt,yshift=-3pt] {\scriptsize $v$} (s6);
        \draw[->] (s4) -- node [left,xshift=-2pt,yshift=-3pt] {\scriptsize $1-0.5\cdot v^2$} (s6);

    \draw[->] (s2) -- node [auto] {\scriptsize $v$} (s3);
        \draw[->] (s4) -- node [auto,yshift=-1.25pt] {\scriptsize $0.5\cdot v^2$} (s5);
        \draw[->] (s5) -- node [right,xshift=1pt,yshift=-5pt] {\scriptsize $1-v$} (s3);

   \draw(s3) edge[loop right] node [right] {\scriptsize $1$} (s3);
      \draw(s6) edge[loop right] node [right] {\scriptsize $1$} (s6);
   \draw(s7) edge[loop right] node [right] {\scriptsize $1$} (s7);
\end{tikzpicture}

}
\caption{A pMC/upMC with parameter $v$.}
\label{fig:example_uMC}
\end{figure}

\begin{example}
Consider the pMC in \autoref{fig:example_uMC} with parameter $V=\{v\}$, initial state $s_0$, and target set $T = \{s_3\}$ (used later). 
Transitions are annotated with polynomials over the parameter~$v$.
\end{example}

\noindent A pMDP $\mdp$ is a \emph{Markov decision process (MDP)} if the transition function yields \emph{well-defined} probability distributions, that is, $\probmdp \colon S \times \Act \times S \rightarrow [0,1]$ and $\sum_{s'\in S}\probmdp(s,\act,s') = 1$ for all $s \in S$ and $\act \in \ActS(s)$.
We denote the \emph{parameter space of $\mdp$} by $\paramspace[\mdp]$, which consists of functions $V \rightarrow \RR$ that map parameters to concrete values.
Applying an \emph{instantiation} $u \in \paramspace[\pmdp]$ to a pMDP $\mdp$ yields the \emph{instantiated} MDP $\mdp[u]$ by replacing each $f\in\mathbb{Q}[\Paramvar]$ in $\mdp$ by $f[u]$. 
An instantiation $u$ is \emph{well-defined} for $\mdp$ if the resulting model $\mdp[u]$ is an MDP.
In the remainder, we assume that all parameter instantiations in $\paramspace[\mdp]$ yield well-defined MDPs.
We call $u$ \emph{graph-preserving} if for all $s,s'\in S$ and $\act\in \Act$ it holds that $\probmdp(s,\act,s')\neq 0 \Rightarrow \probmdp(s,\act,s')[u]\in(0,1]$.

\begin{assume}
    We consider only parameter instantiations for upMDPs that are graph-preserving.
\end{assume}

\noindent
To define measures on MDPs, nondeterministic choices are resolved by a \emph{strategy} $\sched\colon S\rightarrow\Act$ with $\sched(s) \in \ActS(s)$.
The set of all strategies over $\mdp$ is $\Sched^\mdp$.
For the specifications we consider in this paper, \emph{memoryless deterministic strategies} are sufficient~\cite{BK08}.
Applying a strategy $\sched$ to an MDP $\mdp$ yields an \emph{induced} Markov chain (MC) $\mdp[\sched]$ where all nondeterminism is resolved.

\paragraph{Measures.}
For an MC $\dtmc$, the \emph{reachability probability} $\prob_\dtmc(\lozenge T)$ describes the (time unbounded) reachability probability of reaching a set $T\subseteq S$ of target states from the initial state $\sinit$~\cite{BK08}. Similar definitions can be given for the step-bounded reachability probability of reaching a set $T$ from the initial state within $k$ steps, and -- given rewards for every state -- the expected rewards accumulated until reaching the target states or the long-run average, and so forth. 

For an MDP $\mdp$, these measures are typically lifted.
The \emph{maximum reachability probability} $\prob_\mdp^\textrm{max}(\lozenge T)$ is the maximum reachability probability in all induced Markov chains (for all strategies $\sched\in\Sched^\mdp$), i.e., $\prob_\mdp^\textrm{max}(\lozenge T) = \max_\sched \prob_{\mdp[\sched]}(\lozenge T)$. Similar definitions hold for the minimums and the other measures described above.
Our approach is directly applicable to more general measures, e.g., measures on paths described by LTL properties~\cite{Pnu77}.

\paragraph{Specifications.}
A specification $\varphi$ combines a measure, a threshold $\lambda$, and a comparison operator from $\{<,\leq,\geq,>\}$.
For example, the specification $\varphi=\reachPropgT = (\prob_\mdp^\textrm{max}(\lozenge T), \leq\lambda)$ specifies that the maximal reachability probability $\prob_\mdp^\textrm{max}(\lozenge T)$  is at most $\lambda$ for the MDP $\mdp$.
If this statement is true for $\mdp$, we say that $\mdp$ satisfies the specification, written as $\mdp \models \varphi$.
For an MC $\dtmc$, $\reachPropgT$ is defined for the measure $\prob_\dtmc(\lozenge T, \leq\lambda)$.

\subsection{Solution Functions}
Recall that every parameter instantiation $u \in \paramspace[\mdp]$ for pMDP $\pmdp$ induces a concrete MDP $\pmdp[u]$. 
For this MDP, we may then compute any of the measures described above.
We exploit this relationship to create a direct mapping from parameter instantiations to real values.

\begin{definition}[Solution Function]
    \label{definition:valueFunction}
    A \emph{solution function} $\sol_\pmdp \colon \paramspace[\pmdp] \to \R$ for pMDP $\pmdp$ is a  function that maps a parameter instantiation $u \in \paramspace[\pmdp]$ to a value $\sol_\pmdp(u)$, called the \emph{solution of $u$}.
\end{definition}

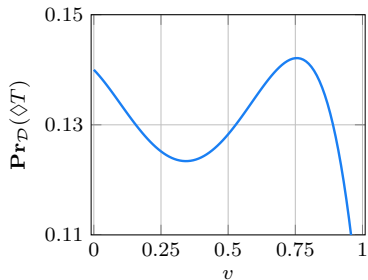
\begin{figure}[t]
\centering
\scalebox{0.9}{

\begin{tikzpicture}[baseline]
\definecolor{color1}{rgb}{0.1,0.498039215686275,0.9549019607843137}
\begin{axis}[width=2.2in,
             height=1.9in,
             ymin=0.11,
             ymax=0.15,
             xmin=-0.01,
             xmax=1.01,
             xmajorgrids,
             ymajorgrids,
             xlabel={$v$},
             ylabel={$\prob_\dtmc(\lozenge T)$},
             xtick={0, 0.25, 0.5, 0.75, 1},
             ytick={0.11, 0.13, 0.15},
             clip=true]

\addplot [color=color1,line width=1pt,domain=0:1, samples=101,unbounded coords=jump, name path=C]{0.5*x^3*(1-x) + 0.1*x^6*(1-x) +  0.1*x*(1-x)^3 +  0.1*(1-x)^2 + 0.04 + 0.05*x};
\end{axis}
\end{tikzpicture}}

\caption{The solution function for the reachability probability of $s_3$ for the pMC in \autoref{fig:example_uMC}.}
\label{fig:example_solfunc}

\end{figure}

\noindent 
In particular, we are interested in solution functions that map a parameter instantiation to the solution of computing a particular measure.\footnote{Notice that we later assume the existence of the integral in \autoref{definition:satprob}, which excludes some esoteric functions.} 
For instance, we may consider a solution function $\sol_\pmdp$ that maps a parameter instantiation $u$ to the probability $\prob_{\mdp[u]}^\textrm{min}(\lozenge T)$. 
In that case, we say that $u$ has solution $\prob_{\mdp[u]}^\textrm{min}(\lozenge T)$. 
\autoref{fig:example_solfunc} depicts a solution function for the reachability probability $\prob_{\mdp}(\lozenge T)$ in the pMC from \autoref{fig:example_uMC}.

Solution functions for parametric models with reachability and expected rewards measures are well-studied, in particular their computation~\cite{Daw04,param_sttt,dehnert-et-al-cav-2015}, but also some of their properties~\cite{winkler2019complexity,DBLP:journals/iandc/BaierHHJKK20}.
Already for pMCs, these functions are typically infeasible to compute. 
In the context of this paper, the important idea is that we can determine the size of the region where this function exceeds a threshold by sampling, as explained next.

\subsection{Uncertain Parametric MDPs}

We now introduce the setting studied in this paper. 
Specifically, we use pMDPs whose parameters define the uncertainty in the transition probabilities of an MDP.
We add another layer of uncertainty, where each parameter follows a probability distribution.
For example, referring back to the UAV example in \autoref{sec:Example}, each weather condition has a certain probability, and every condition leads to a certain parameter instantiation.
Importantly, the probability distribution of the parameters is assumed to be \emph{unknown}, and we just assume that we are able to sample this distribution.

\begin{definition}[upMDP]
	An \emph{uncertain pMDP} (upMDP) $\vmdp$ is a tuple $\vmdp = \left(\pmdp, \probdist \right)$ with $\pmdp$ a pMDP, and $\probdist$ a probability distribution over the parameter space $\paramspace[\pmdp]$.
	If $\pmdp$ is a pMC, then we call $\vmdp$ a upMC.
\end{definition}

\noindent
Intuitively, a upMDP is a pMDP with an associated distribution over possible  parameter instantiations.
That is, sampling from $\paramspace[\pmdp]$ according to $\probdist$ yields concrete MDPs $\pmdp[u]$ with instantiations $u\in\paramspace[\pmdp]$ (and $\probdist(u)>0$).

\begin{definition}[Satisfaction Probability]
\label{definition:satprob}
Let $\vmdp = \left(\pmdp, \probdist\right)$ be a upMDP and $\varphi$ a specification.
	The (weighted) \emph{satisfaction probability of $\varphi$} in $\vmdp$ is  
	\[ \satprob= \int_{\paramspace[\pmdp]}   I_\varphi(u)\; d\,\probdist(u)\]
	with $u\in \paramspace[\pmdp]$ and $I_\varphi \colon \paramspace[\pmdp] \rightarrow \{0,1\}$ is the indicator for $\varphi$, i.e.\ $I_\varphi(u) = 1$ iff $\pmdp[u] \models \varphi$. 
\end{definition}

\noindent
Note that $I_\varphi$ is measurable for all specifications mentioned in this paper, as it partitions  $\paramspace[\pmdp]$ into a finite union of semi-algebraic sets~\cite{Book_Basu_RAAlgorithms,winkler2019complexity}.  
Moreover, we have that $\satprob \in [0,1]$ and

\begin{align}
\label{eq:satprobsumstoone}
    \satprob + \falseprob= 1.
    \end{align}

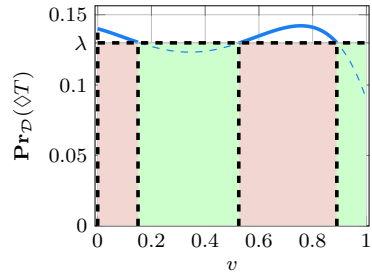
\begin{figure}[t]
\centering
\scalebox{0.9}{

\begin{tikzpicture}[baseline]
\definecolor{color1}{rgb}{0.1,0.498039215686275,0.9549019607843137}
\begin{axis}[width=2.2in,height=1.9in,ymin=-0.0001,xmin=-0.01,xmax=1.01,xmajorgrids,ymajorgrids,xlabel={$v$},ylabel={$\prob_\dtmc(\lozenge T)$},ytick={0,0.05,0.1,0.13,0.15},yticklabels={0,0.05,0.1,$\lambda$,0.15},clip=false]
  
\addplot [color=color1,line width=0.5pt,dashed,domain=0:1, samples=101,unbounded coords=jump, name path=C]{0.5*x^3*(1-x) + 0.1*x^6*(1-x) +  0.1*x*(1-x)^3 +  0.1*(1-x)^2 + 0.04 + 0.05*x};
\addplot [color=color1,line width=1.5pt,domain=0:0.15, samples=101,unbounded coords=jump, name path=CC]{0.5*x^3*(1-x) + 0.1*x^6*(1-x) +  0.1*x*(1-x)^3 +  0.1*(1-x)^2 + 0.04 + 0.05*x};
\addplot [color=gray!50!white,line width=0pt,domain=0:0.15, samples=10,unbounded coords=jump, name path=DD]{0};
\addplot [color=gray!50!white,line width=0pt,domain=0.15:0.525, samples=10,unbounded coords=jump, name path=EE]{0};
\addplot [color=color1,line width=1.5pt,domain=0.525:0.89, samples=101,unbounded coords=jump, name path=CCC]{0.5*x^3*(1-x) + 0.1*x^6*(1-x) +  0.1*x*(1-x)^3 +  0.1*(1-x)^2 + 0.04 + 0.05*x};
\addplot [color=gray!50!white,line width=0pt,domain=0.525:0.89, samples=10,unbounded coords=jump, name path=DDD]{0};
\addplot [color=gray!50!white,line width=0pt,domain=0.89:1, samples=10,unbounded coords=jump, name path=EEE]{0};
\addplot [color=gray!50!white,line width=0pt,domain=0:0.15, samples=10,unbounded coords=jump, name path=DDA]{0.13};
\addplot [color=gray!50!white,line width=0pt,domain=0.15:0.525, samples=10,unbounded coords=jump, name path=EEA]{0.13};
\addplot [color=gray!50!white,line width=0pt,domain=0.525:0.89, samples=10,unbounded coords=jump, name path=DDDA]{0.13};
\addplot [color=gray!50!white,line width=0pt,domain=0.89:1, samples=10,unbounded coords=jump, name path=EEEA]{0.13};
     \addplot[red!20!white,domain=0:0.15] fill between[of=DD and DDA];
    \addplot[red!20!white,domain=0.525:0.89]  fill between[of=DDD and DDDA];
         \addplot[green!20!white,domain=0.15:0.525] fill between[of=EE and EEA];
    \addplot[green!20!white,domain=0.89:1]  fill between[of=EEE and EEEA];
    \draw [color=black,dashed,line width=1.5pt](axis cs:0,0.0) -- node[left]{} (axis cs:0,0.14);
    \draw [color=black,dashed,line width=1.5pt](axis cs:0.15,0.0) -- node[left]{} (axis cs:0.15,0.13);
    \draw [color=black,dashed,line width=1.5pt](axis cs:0.525,0.0) -- node[left]{} (axis cs:0.525,0.13);
    \draw [color=black,dashed,line width=1.5pt](axis cs:0.89,0.0) -- node[left]{} (axis cs:0.89,0.13);
    \draw [color=black,dashed,line width=1.5pt](axis cs:0,0.13) -- node[left]{} (axis cs:1.0,0.13);
\end{axis}
\end{tikzpicture}}

\caption{The probability of satisfying the reachability specification $\varphi=\reachPropgT$ for the upMC in \autoref{fig:example_uMC}, versus the value of the parameter $v$. Intervals that satisfy $\varphi$ are green, intervals that violate $\varphi$ are red.}
\label{fig:example_satisfactionProb}

\end{figure}

\begin{example}
\label{ex:running-umc}
We expand the pMC in \autoref{fig:example_uMC} toward a upMC with a uniform distribution for the parameter $v$ over the interval $[0,1]$. 
In \autoref{fig:example_satisfactionProb}, we again plot the solution function for the reachability probability in the pMC from \autoref{fig:example_uMC}, which was also shown in \autoref{fig:example_solfunc}.
Additionally, we compare this probability against a threshold $\lambda = 0.13$ with comparison operator $\leq$, and we plot the satisfying region and its complementary as green and red regions.
The satisfying region is given by the union of the intervals $\left[0.13, 0.525\right]$ and $\left[0.89,1.0\right]$, and the satisfaction probability $\satreachprob$ is $0.395+0.11=0.505$.
\end{example} 

%% file: sections/4_problem.tex
\section{Problem Statement}
\label{sec:Problem}

Let us now formalize the problem of interest.
We aim to compute the satisfaction probability of the parameter space for a specification $\varphi$ on a upMDP.
Equivalently, we thus seek the probability that a randomly sampled instantiation $u$ from the parameter space $\paramspace[\pmdp]$ induces an MDP $\pmdp[u]$ which satisfies $\varphi$.
Formally: given a upMDP $\vmdp=\left(\pmdp,  \probdist\right)$, and a specification $\varphi$, compute the satisfaction probability $\satprob$. 
We \emph{approximate} this probability by sampling the parameters.
Such an approach cannot be precise and correct in all cases, because we only have a finite number of samples at our disposal.
Instead, we provide the following \emph{probably approximately correct} (PAC) style formulation~\cite{DBLP:conf/aaai/Haussler90}, meaning that we compute a \emph{lower bound} on the satisfaction probability that is correct with \emph{high confidence}:

\begin{mdframed}[backgroundcolor=gray!30]
    \textbf{Formal problem 1.} \, Given a upMDP $\vmdp=\left(\pmdp,  \probdist\right)$, a specification $\varphi$, and a confidence probability $\confidence \in (0,1)$, compute a lower bound $\eta$ on the satisfaction probability, such that $\satreachprob \geq \eta$ holds with a confidence probability of at least $\confidence$.
\end{mdframed}

\noindent 
Intuitively, given a confidence probability $\beta$ close below one, we obtain $\eta$ as a \emph{high-confidence lower bound on the satisfaction probability} $\satreachprob$.

\begin{remark}
    We can also compute an upper bound on the satisfaction probability by exploiting \eqref{eq:satprobsumstoone} and computing a lower bound for the negated specification $\neg \varphi$.
\end{remark}

Furthermore, as is typical in PAC settings, if a specific value for $\eta$ is desired, we are also able to compute the \emph{confidence that $\eta$ is indeed a lower bound on the satisfaction probability}.
We will exploit both directions, with either given $\beta$ or $\eta$, in the numerical examples in \autoref{sec:numerical_examples}.
We illustrate our formal problem by continuing our examples on the upMC in \autoref{fig:example_uMC} and the UAV.

\begin{example}
\label{ex:running-prob}
  Let us reconsider the upMC from \autoref{ex:running-umc} with $\varphi$ and  satisfaction probability $\satreachprob = 0.505$. 
  Assume we do not yet know this probability. 
The problem statement then asks how to compute an $\eta$ such that with high confidence $\beta$, say $0.99$, $\satreachprob \geq \eta$.
\end{example}

\begin{example}
    For the UAV motion planning example introduced in \autoref{sec:Example}, consider the question \textquotedblleft \emph{What is a lower bound on the probability that on a given day, there exists a strategy for the UAV to successfully complete the mission?}\textquotedblright \,
    Our specification $\varphi$ for successfully completing a mission could then be that the maximal reachability probability to a target state is above $0.99$, or that the expected travel time is below $12$ hours. 
    For any such $\varphi$, assume that we want to answer the question above with confidence of $\beta=0.9$.
    The resulting lower bound on the satisfaction probability could be, e.g., $\eta = 0.81$.
    This means that with a confidence probability of $\confidence = 0.9$, the actual satisfaction probability is indeed at least $\eta = 0.81$. 
    If we change the confidence $\beta$ to $0.99$, the obtained lower bound may reduce to $\eta = 0.78$. 
    Intuitively, the more confidence we want to have, the lower the lower bound.
\end{example}

%% file: sections/5_robust.tex
\section{Computing the Satisfaction Probability}
\label{sec:scenario:closetoone}
\begin{figure}[t!]
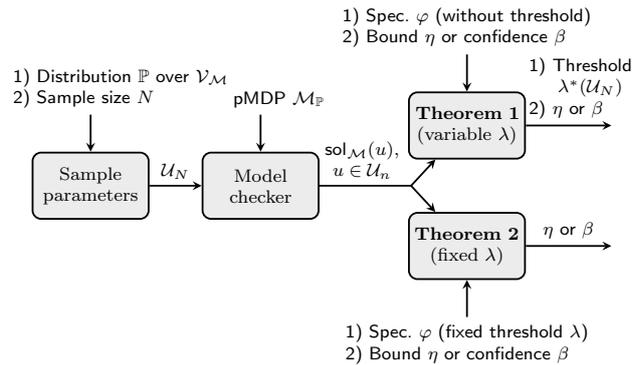

	\centering
	\include{tikz/algorithm}
	\caption{
	    Schematic overview of our approach.
	    After obtaining the solutions, we choose to apply \autoref{theorem:reach_bound} (outputting a specification threshold $\lambda^*(\mathcal{U}_N)$ depending on $\mathcal{U}_N$) or \autoref{theorem:reach_bound_relax} (inputting a fixed threshold $\lambda$).
	}
	\label{fig:Approach}
\end{figure}

In this section, we introduce our approach for solving the problem presented in \autoref{sec:Problem}. 
We focus on a practical overview of our approach in this section, while postponing technical details and the derivation of our main results (Theorems~\ref{theorem:reach_bound} and \ref{theorem:reach_bound_relax}) to \autoref{sec:derivations}.
First, in \autoref{subsec:Overview}, we fix some notation for our concrete setup. 
In particular, we discuss how we obtain solutions $\sol_\pmdp(u)$ by sampling parameter instantiations $u$ from $\paramspace[\mdp]$.
In \autoref{subsec:scenario:noviolations}, we then first address a simpler yet related problem in which we let the specification $\varphi$ depend on the set of sampled solutions at hand.
In \autoref{subsec:scenario:violations}, we return to our original problem statement: we introduce our approach in which we keep the specification fixed.
An algorithmic overview of both of these methods is shown in \autoref{fig:Approach}. 
Finally, we discuss the quality of the obtained lower bounds in \autoref{subsec:scenario:quality}

\subsection{Obtaining Solutions from Parameter Samples}
\label{subsec:Overview}

We describe how we obtain solutions by sampling from the parameter space.
Specifically, we define the set $\mathcal{U}_N = \{ u_1, u_2, \ldots, u_N \}$ as the outcome of sampling $N$ parameter instantiations from $\paramspace[\mdp]$ according to the probability distribution $\PP$.
Recall that we assume that these samples are independent and identically distributed.
Thus, the set $\mathcal{U}_N$ of $N$ samples is a random element drawn according to the product probability $\probdist^N = \probdist \times \cdots \times \probdist$ ($N$ times) over the parameter probability distribution $\probdist$.
For each sample $u \in \mathcal{U}_N$, we compute the resulting solution $\sol_{\pmdp}(u)$, as shown in the following example.

\begin{figure}[t]
    \centering
        \input{tikz/scenarioApproach_visual}
    \caption{A set of $N=10$ parameter instantiations $\mathcal{U}_N = \{ u_{1}, \ldots, u_{10} \}$ (shown as red crosses) for the upMC in \autoref{fig:example_uMC} and the solutions $\sol_{\pmdp}(u)$.
    In \autoref{fig:scenarioApproach_visual_left}, the specification threshold $\lambda^*(\mathcal{U}_N)$ is chosen after observing the solutions such that all samples are satisfying; \autoref{fig:scenarioApproach_visual_right} uses a fixed threshold $\lambda$ and has two violating samples.
    }
    
    \label{fig:scenarioApproach_visual}
\end{figure}
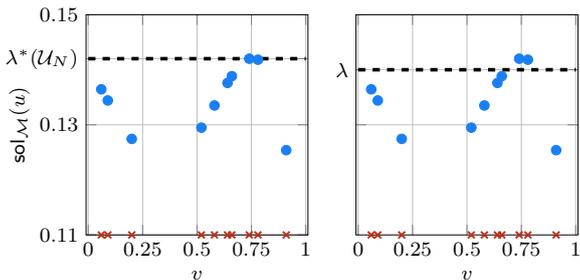

\begin{example}
\label{ex:Samples}
    We continue from \autoref{ex:running-prob} on the upMC in \autoref{fig:example_uMC}.
    We sample $N=10$ parameters from the (uniform) probability distribution of this upMC, which are shown in red on the x-axis in \autoref{fig:scenarioApproach_visual}. 
    The resulting solutions $\sol_{\pmdp}(u_1), \ldots, \sol_{\pmdp}(u_{10})$ are depicted as the blue points.
    As expected, these points indeed lie on the solution function curve shown in \autoref{fig:example_solfunc}.
    In \autoref{fig:scenarioApproach_visual_left} and \ref{fig:scenarioApproach_visual_right}, we check these solutions against a specification in two different ways:
    (1)~In \autoref{fig:scenarioApproach_visual_left}, we first observe the solutions and then devise a threshold for the given measure, such that these samples \emph{all satisfy the resulting specification}.
    That is, the threshold, denoted by $\lambda^*(\mathcal{U}_N)$, depends on the solutions at hand, such that $\pmdp[u] \models \varphi$ for all $u \in \mathcal{U}_N$.
    For this example, the specification is $\reachPropgTvary$, and the tightest threshold satisfying this condition is $\lambda^*(\mathcal{U}_N) = \max_{u \in \mathcal{U}_N} \sol_{\pmdp}(u)$.
    (2)~In \autoref{fig:scenarioApproach_visual_right}, we fix the specification with its threshold first, and then evaluate the number of samples satisfying the specification.
    This may lead to samples violating the specification (e.g., \autoref{fig:scenarioApproach_visual_right} has two violating samples).
\end{example}

In both cases, we can partition $\mathcal{U}_N$ into disjoint sets of samples that satisfy ($\mathcal{U}_{\satisfying}$) or violate ($\mathcal{U}_{\violating}$) the specification, i.e., $\mathcal{U}_N=\mathcal{U}_{\satisfying}\cup \mathcal{U}_{\violating}$.
Note, that in the first case (\autoref{fig:scenarioApproach_visual_left}), the set of violating samples is empty by construction, i.e., $\mathcal{U}_{\violating} = \varnothing$.
Let $\satisfying = \vert \mathcal{U}_{\satisfying} \vert$ denote the number of satisfying samples and $\violating = \vert \mathcal{U}_{\violating} \vert$ the number of violating samples.

\subsection{Restriction to Satisfying Samples}
\label{subsec:scenario:noviolations}

Before solving the main problem introduced in \autoref{sec:Problem}, we consider a simpler setting to introduce some of the necessary ideas. 
Intuitively, we want to investigate the case where we adapt the specification (or rather the threshold in this specification) such that $\mathcal{U}_N=\mathcal{U}_{\satisfying}$, or equivalently $N_{\varphi} = N$. This simpler setting is shown by  \autoref{fig:scenarioApproach_visual_left}. Here we do not fix a threshold $\lambda$ for the specification $\varphi$ a-priori, but instead derive a threshold $\lambda^*(\mathcal{U}_N)$ from the solutions at hand such that that \emph{all samples are satisfying}, i.e. we ensure that
\begin{align}
    \label{eq:AssumptionNonviolating}
    \mdp[u] \models \varphi \text{ for all samples }u \in \mathcal{U}_N.\tag{Assumption A}
\end{align}

\paragraph{Problem.}
We raise the question: \textquotedblleft\emph{What is the probability that, given these $N$ samples and a specification threshold that makes all samples satisfying, the next sampled parameter valuation $u$ (on the x-axis of \autoref{fig:scenarioApproach_visual_left}) with the corresponding solution $\sol_{\pmdp}(u)$ will also satisfy this specification?}\textquotedblright \,
This probability is similar to the satisfaction probability $\satreachprob$ in \autoref{definition:satprob}, but the threshold of specification $\varphi$ is not fixed a-priori.

\paragraph{Result.}
Using \autoref{theorem:reach_bound}\footnote{We derive this theorem using \autoref{lemma:reach_bound}, which is provided later on in \autoref{subsec:scenario:Basic}}, we compute a lower bound $\eta$ on this satisfaction probability that  holds with a user-specified confidence probability $\beta$:

\begin{theorem}
    \label{theorem:reach_bound}
    Let upMDP $\vmdp$ and the set $\mathcal{U}_N$ of $N \geq 1$ sampled parameters.
    For any set $\mathcal{U}_N$, choose threshold $\lambda^*(\mathcal{U}_N)$ of specification $\varphi$ such that $\mathcal{U}_N=\mathcal{U}_{\satisfying}$, and fix a confidence probability $\confidence \in (0,1)$.
    Then, it holds that
    \begin{equation}
        \label{eq:reach_bound} 
        \probdist^N \Big\{ \satreachprob \geq (1-\confidence)^{\frac{1}{N}} \Big\} \geq \beta.
    \end{equation}
\end{theorem}

\noindent
Applying \autoref{theorem:reach_bound} to the solutions in \autoref{fig:scenarioApproach_visual_left}, we compute that the satisfaction probability $\satreachprob$ with respect to specification $\varphi$ with threshold $\lambda^*(\mathcal{U}_N) = 0.142$ is bounded from below by $\eta = 0.794$ (with a confidence probability of at least $\beta = 0.9$) and by $\eta = 0.631$ (with a confidence of at least $\beta = 0.99$).

\paragraph{Sample complexity.}
More generally, \autoref{theorem:reach_bound} asserts that with a probability of at least $\confidence$, the next sampled parameter from $\paramspace[\mdp]$ will satisfy the specification (with sample-dependent threshold $\lambda^*(\mathcal{U}_N)$) with a probability of at least $(1-\beta)^{\frac{1}{N}}$.
Thus, the satisfaction probability is lower bounded by $\eta = (1-\beta)^{\frac{1}{N}}$ with high confidence, given that $\confidence$ is close to one.
This high confidence is easily achieved for a sufficiently large number of samples $N$, as seen from the following corollary.

\begin{coro}
    \label{coro:SampleComplexity}
    The sample size $N$ necessary to obtain a desired lower bound $\eta \in (0,1)$ on the satisfaction probability with at least a confidence of $\confidence \in (0,1)$ is
    \begin{equation}
        N = \text{ceil} \Big( \frac{\log (1-\confidence)}{\log \eta} \Big),
    \end{equation}
    where the function $\text{ceil}(x)$ rounds its argument $x \in \R$ upwards to the nearest integer.
\end{coro}

\noindent 
\autoref{coro:SampleComplexity} states that the sample size $N$ is logarithmic in the confidence probability $\beta$.
Thus, a significant improvement in $\beta$ (i.e. closer to one) only requires a marginal increase in $N$.
Similarly, increasing the sample size $N$ improves the lower bound on the satisfaction probability $\eta$.
For example, when increasing the number of samples in \autoref{fig:scenarioApproach_visual_left} to $N=100$ (note that we still assume that $\pmdp[u] \models \varphi$ for all $u \in \mathcal{U}_N$), \autoref{theorem:reach_bound} concludes that the satisfaction probability is lower bounded by $0.977$ (with a confidence of at least $\beta = 0.9$) and by $0.955$ (with a confidence of at least $\beta = 0.99$).
Next, consider the extreme case, where $\confidence$ is infinitely close to one.
We observe from Corollary~\ref{coro:SampleComplexity} that such a confidence probability can only be obtained for $N = \infty$.
Intuitively, this observation makes sense: we can only be absolutely certain of our lower bound on the satisfaction probability, if we have based this estimate on infinitely many samples.
In practice, our sample set is finite, and a typical confidence probability may be $\confidence = 1 - 10^{-3}$.

\subsection{Treatment of Violating Samples}
\label{subsec:scenario:violations}

We return to a setting with a fixed threshold $\lambda$ and possible violating samples. 
In other words, we violate \eqref{eq:AssumptionNonviolating} that $\mdp[u] \models \varphi$ for all samples $u \in \mathcal{U}_N$.
Consider again \autoref{fig:scenarioApproach_visual_right}, where for some of the samples $u \in \mathcal{U}_N$, the value $\sol_{\pmdp}(u)$ exceeds $\lambda$, so $\mathcal{U}_N\neq\mathcal{U}_{\satisfying}$, and  \autoref{theorem:reach_bound} does not apply. 
Instead, we state \autoref{theorem:reach_bound_relax}\footnote{We derive this theorem using \autoref{lemma:reach_bound_relax}, which is provided later on in \autoref{subsec:scenario:ViolatingSamples}} as a generalization that uses a fixed threshold $\lambda$ and is also applicable in the presence of violating samples:

\begin{theorem}
    \label{theorem:reach_bound_relax}
    Let upMDP $\vmdp$ and the set $\mathcal{U}_N$ of $N \geq 1$ sampled parameters.
    Choose a confidence probability $\confidence \in (0,1)$.
    Then, it holds that
    \begin{equation}
        \probdist^N \Big\{ 
            \satreachprob \, \geq \, t^*(\violating)
        \Big\} \geq \confidence,
        \label{eq:reach_bound_relax_2}
    \end{equation}
    where $t^*(N) = 0$ for $\violating = N$, and for $k = 0, \ldots, N-1$, $t^*(k)$ is the solution of
    \begin{equation}
        \frac{1 - \beta}{N} = \sum_{i=0}^{k} \binom Ni (1-t)^i t^{N-i}.
        \label{eq:reach_bound_relax_1}
    \end{equation}
\end{theorem}

\noindent
\autoref{theorem:reach_bound_relax} solves the formal problem stated in \autoref{sec:Problem}.
Recall that $\violating$ denotes the number of samples whose value $\sol_{\pmdp}(u)$ violates the specification $\varphi$. 
Applying \autoref{theorem:reach_bound_relax} to the solutions in \autoref{fig:scenarioApproach_visual_right} (with $\violating=2$), we conclude that the satisfaction probability is bounded from below by $0.388$ (with a confidence of at least $\beta = 0.9$) and by $0.282$ (with $\beta = 0.99$).
When we increase the number of samples to $N=100$ and assume that $\violating = 20$, these results improve to the lower bounds $0.654$ (with $\beta = 0.9$) and $0.622$ (with $\beta = 0.99$).
We note that the intuition in \autoref{coro:SampleComplexity} about the relationships between the sample size $N$, lower bound $\eta$, and the confidence probability $\beta$ also holds for \autoref{theorem:reach_bound_relax}.

\begin{figure}[t!]
    \centering
    \include{tikz/scenario-plots/fixedN}
     \caption{Lower bounds on the satisfaction probability as computed by \autoref{theorem:reach_bound} (shown as points at $\violating=0$) and \autoref{theorem:reach_bound_relax} (lines for different $\violating = 0,\ldots,N$).}   
    \label{fig:scenario-fixedN}
\end{figure}

\begin{figure}[t!]
    \centering
    \include{tikz/scenario-plots/fixedFraction}
     \caption{Bounds on the satisfaction probability from \autoref{theorem:reach_bound_relax} for fixed fractions $\violating / N$ of violated samples.}    
    \label{fig:scenario-fixedFraction}
\end{figure}

\subsection{Quality of the Obtained Lower Bounds}
\label{subsec:scenario:quality}
\autoref{fig:scenario-fixedN} illustrates how the number of violating samples, $\violating$, influences the quality of the lower bound on the satisfaction probability. 
The points at $\violating=0$ are the bounds returned by \autoref{theorem:reach_bound}, while the lines correspond to \autoref{theorem:reach_bound_relax}.
Intuitively, the lower bound on the satisfaction probability computed by \autoref{theorem:reach_bound_relax} decreases with an increased number of violating samples.
Moreover, \autoref{theorem:reach_bound} yields a better lower bound than \autoref{theorem:reach_bound_relax} (points versus the lines in \autoref{fig:scenario-fixedN}), at the cost of not using a fixed threshold on the specification, and not being able to deal with violating samples.

In \autoref{fig:scenario-fixedFraction}, we fix the fraction of violating samples $\nicefrac{\violating}{N}$ and plot the lower bounds on the satisfaction probability obtained using \autoref{theorem:reach_bound_relax} for different values of $N$ and $\beta$.
Note that the lower bounds grow toward the fraction of violation for increased sample sizes.
As also shown with \autoref{coro:SampleComplexity}, the confidence probability $\beta$ only has a marginal effect on the obtained lower bounds.

Finally, we make the following remark with respect to the sample complexity of Theorems~\ref{theorem:reach_bound} and~\ref{theorem:reach_bound_relax}.

\begin{remark}[Independence to model size]\label{remark:independent}
	The number of samples needed to obtain a certain confidence probability in Theorems~\ref{theorem:reach_bound} and~\ref{theorem:reach_bound_relax} is independent of the number of states, transitions, or parameters of the upMDP.\footnote{Despite this independence, note that the time to compute solutions via model checking still depends on the number of states and transitions of the instantiated MDP.}
\end{remark}

%% file: tikz/algorithm.tex
\usetikzlibrary{calc}
\usetikzlibrary{arrows.meta}
\usetikzlibrary{positioning}
\usetikzlibrary{fit}
\usetikzlibrary{shapes}

\tikzstyle{node} = [rectangle, rounded corners, minimum width=1.7cm, text width=1.7cm, minimum height=1.1cm, text centered, draw=black, fill=gray!15]

\resizebox{\linewidth}{!}{%
	\begin{tikzpicture}[node distance=4.2cm,->,>=stealth,line width=0.3mm,auto,
		main node/.style={circle,draw,font=\sffamily\bfseries}]
		
		\newcommand\xshift{-1cm}
		\newcommand\yshift{-2.9cm}
		
		\node (sampler) [node] {Sample \\ parameters};
		\node (checker) [node, right of=sampler, xshift=-1.4cm] {Model \\ checker};
		\node (split) [right of=checker, xshift=-1.6cm] {};
		\node (thm1) [node, right of=checker, xshift=-0.8cm, yshift=1cm] {\textbf{Theorem 1} \\ (variable $\lambda$)};
		\node (thm2) [node, right of=checker, xshift=-0.8cm, yshift=-1cm] {\textbf{Theorem 2} \\ (fixed $\lambda$)};
		
		
		\node (in1) [above of=sampler, yshift=\yshift] {};
		\node (in2) [above of=checker, yshift=\yshift] {};
		\node (in_thm1) [above of=thm1, yshift=\yshift] {};
		\node (in_thm2) [below of=thm2, yshift=-\yshift] {};
		\node (out1) [right of=thm1, xshift=-1.7cm] {};
		\node (out2) [right of=thm2, xshift=-1.7cm] {};
		
		
		\draw [->] (in1) -- (sampler) node [xshift=0.5cm, pos=0.0, above, align=left, font=\sffamily\small] {1) Distribution $\PP$ over $\paramspace[\pmdp]$ \\ 2) Sample size $N$};
		
		\draw [->] (in2) -- (checker) node [xshift=0.3cm, pos=0.0, above, align=left, font=\sffamily\small] {pMDP $\vmdp$};
		
		\draw [->] (in_thm1) -- (thm1) node [pos=0.0, above, align=left, font=\sffamily\small] {1) Spec. $\varphi$ (without threshold) \\ 2) Bound $\eta$ or confidence $\beta$};
		
        \draw [->] (in_thm2) -- (thm2) node [pos=0.0, below, align=left, font=\sffamily\small] {1) Spec. $\varphi$ (fixed threshold $\lambda$) \\ 2) Bound $\eta$ or confidence $\beta$};
		
		
		\draw [->] (sampler) -- (checker) node [pos=0.5, above, align=center, font=\sffamily\small] {$\mathcal{U}_N$};
		
		\draw [-] (checker) -- (split) node [pos=0.5, above, align=center, font=\sffamily\small, xshift=-0.05cm] {$\sol_\pmdp(u),$ \\ $u \in \mathcal{U}_n$};
		
		\draw [->] (split.west) -- (thm1) {};
		\draw [->] (split.west) -- (thm2) {};
		
		\draw [->] (thm1) -- (out1) node [pos=0.64, xshift=-0.0cm, above, align=left, font=\sffamily\small] {1) Threshold \\ $\quad\enskip\lambda^*(\mathcal{U}_N)$ \\ 2) $\eta$ or $\beta$};
		
		\draw [->] (thm2) -- (out2) node [pos=0.5, xshift=-0.0cm, above, align=left, font=\sffamily\small] {$\eta$ or $\beta$};
		
	\end{tikzpicture}
}

%% file: tikz/scenarioApproach_visual.tex
\begin{subfigure}[b]{0.47\linewidth}

\scalebox{0.9}{
\begin{tikzpicture}[baseline]
\definecolor{color1}{rgb}{0.1,0.498039215686275,0.9549019607843137}
\begin{axis}[
    width=1.9in,
    height=1.9in,
    ymin=0.11,
    ymax=0.15,
    xmin=-0.01,
    xmax=1.01,
    xmajorgrids,
    ymajorgrids,
    xlabel={$v$},
    ylabel= {$\sol_\pmdp(u)$}, 
    y label style={at={(axis description cs:-0.22,0.5)}},
    xtick={0,0.25,0.5,0.75,1},
    ytick={0.11, 0.13, 0.142009112744576, 0.15},
    yticklabels={0.11, 0.13, $\lambda^*(\mathcal{U}_N)$, 0.15},
    clip=false]

\addplot [only marks, samples at={0.58, 0.91, 0.66, 0.09, 0.64, 0.06, 0.74, 0.52, 0.78, 0.2}, mark options={draw=color1, fill=color1},]{0.5*x^3*(1-x) + 0.1*x^6*(1-x) +  0.1*x*(1-x)^3 +  0.1*(1-x)^2 + 0.04 + 0.05*x};

\addplot[color=red, mark=x, only marks, thick] coordinates {%
		(0.58, 0.11)
		(0.91, 0.11)
		(0.66, 0.11)
		(0.09, 0.11)
		(0.64, 0.11)
		(0.06, 0.11)
		(0.74, 0.11)
		(0.52, 0.11)
		(0.78, 0.11)
		(0.20, 0.11)
	};

\draw [color=black,dashed,line width=1.5pt](axis cs:0,0.142009112744576) -- (axis cs:1,0.142009112744576);

\end{axis}
\end{tikzpicture}}

\caption{All samples $u \in \mathcal{U}_N$ correspond to $\sol_\pmdp(u) \leq \lambda^*(\mathcal{U}_N)$.}
\label{fig:scenarioApproach_visual_left}

\end{subfigure}
\hfill
\begin{subfigure}[b]{0.47\linewidth}
\scalebox{0.9}{
\begin{tikzpicture}[baseline]
\definecolor{color1}{rgb}{0.1,0.498039215686275,0.9549019607843137}
\begin{axis}[
    width=1.9in,
    height=1.9in,
    ymin=0.11,
    ymax=0.15,
    xmin=-0.01,
    xmax=1.01,
    xmajorgrids,
    ymajorgrids,
    xlabel={$v$},
    xtick={0, 0.25, 0.5, 0.75, 1},
    ytick={0.1, 0.13, 0.14, 0.15},
    yticklabels={,,$\lambda$,},
    clip=false]

\addplot [only marks, samples at={0.58, 0.91, 0.66, 0.09, 0.64, 0.06, 0.74, 0.52, 0.78, 0.2}, mark options={draw=color1, fill=color1},]{0.5*x^3*(1-x) + 0.1*x^6*(1-x) +  0.1*x*(1-x)^3 +  0.1*(1-x)^2 + 0.04 + 0.05*x};

\addplot[color=red, mark=x, only marks, thick] coordinates {%
		(0.58, 0.11)
		(0.91, 0.11)
		(0.66, 0.11)
		(0.09, 0.11)
		(0.64, 0.11)
		(0.06, 0.11)
		(0.74, 0.11)
		(0.52, 0.11)
		(0.78, 0.11)
		(0.20, 0.11)
	};

\draw [color=black,dashed,line width=1.5pt](axis cs:0,0.14) -- (axis cs:1,0.14);

\end{axis}
\end{tikzpicture}}%

\caption{Two samples $u \in \mathcal{U}$ correspond to $\sol_\pmdp(u) > \lambda$.}
\label{fig:scenarioApproach_visual_right}

\end{subfigure}

%% file: tikz/scenario-plots/fixedN.tex
\begin{subfigure}[b]{\linewidth}

    \begin{tikzpicture}
      \begin{axis}[
          width=\linewidth,
          height=4.5cm,
          ymajorgrids,
          grid style={dashed,gray!30},
          xlabel=Number of violating samples ($\violating$),
          ylabel=Lower bound ($\eta$),
          x label style={at={(axis description cs:0.4,-0.2)}},
          x tick label style={rotate=90,anchor=east},
          xmin=-0.1,
          xmax=10,
          ymin=0,
          ymax=1,
          every axis plot/.append style={line width=1pt},
          legend cell align={left},
          legend columns=1,
          legend pos=north east,
        ]
        
        \addplot[mark=none, color=MidnightBlue, mark size=1.3pt] table[x=k, y=b0.9, col sep=semicolon] {tikz/scenario-plots/eta_risk_fixedN=10.csv};
        
        \addplot[mark=none, color=BurntOrange, mark size=1.3pt] table[x=k, y=b0.99, col sep=semicolon] {tikz/scenario-plots/eta_risk_fixedN=10.csv};
        
        \addplot[mark=none, color=OliveGreen, mark size=1.3pt] table[x=k, y=b0.999, col sep=semicolon] {tikz/scenario-plots/eta_risk_fixedN=10.csv};
        
        \node[circle,fill,inner sep=2pt,fill=MidnightBlue] at (axis cs:0,0.79432823) {};
        \node[circle,fill,inner sep=2pt,fill=BurntOrange] at (axis cs:0,0.63095734) {};
        \node[circle,fill,inner sep=2pt,fill=OliveGreen] at (axis cs:0,0.50118723) {};
        
        \legend{{$\beta=0.9$},{$\beta=0.99$},{$\beta=0.999$}}
      \end{axis}
    \end{tikzpicture}

\caption{Number of samples $N=10$.}
\label{fig:scenario_fixedN=10}

\end{subfigure}
\hfill
\begin{subfigure}[b]{\linewidth}
    \begin{tikzpicture}
      \begin{axis}[
          width=\linewidth,
          height=4.5cm,
          ymajorgrids,
          grid style={dashed,gray!30},
          xlabel=Number of violating samples ($\violating$),
          ylabel=Lower bound ($\eta$),
          x label style={at={(axis description cs:0.4,-0.2)}},
          x tick label style={rotate=90,anchor=east},
          xmin=-1,
          xmax=100,
          ymin=0,
          ymax=1,
          every axis plot/.append style={line width=1pt},
          legend cell align={left},
          legend columns=1,
          legend pos=north east,
        ]
        
        \addplot[mark=none, color=MidnightBlue, mark size=1.3pt] table[x=k, y=b0.9, col sep=semicolon] {tikz/scenario-plots/eta_risk_fixedN=100.csv};
        
        \addplot[mark=none, color=BurntOrange, mark size=1.3pt] table[x=k, y=b0.99, col sep=semicolon] {tikz/scenario-plots/eta_risk_fixedN=100.csv};
        
        \addplot[mark=none, color=OliveGreen, mark size=1.3pt] table[x=k, y=b0.999, col sep=semicolon] {tikz/scenario-plots/eta_risk_fixedN=100.csv};
        
        \node[circle,fill,inner sep=2pt,fill=MidnightBlue] at (axis cs:0,0.97723722) {};
        \node[circle,fill,inner sep=2pt,fill=BurntOrange] at (axis cs:0,0.95499259) {};
        \node[circle,fill,inner sep=2pt,fill=OliveGreen] at (axis cs:0,0.9332543) {};
        
        \legend{{$\beta=0.9$},{$\beta=0.99$},{$\beta=0.999$}}
      \end{axis}
    \end{tikzpicture}

\caption{Number of samples $N=100$.}
\label{fig:scenario_fixedN=100}

\end{subfigure}

%% file: tikz/scenario-plots/fixedFraction.tex
\begin{subfigure}[b]{\linewidth}

    \begin{tikzpicture}
      \begin{axis}[
          width=\linewidth,
          height=4.5cm,
          ymajorgrids,
          grid style={dashed,gray!30},
          xlabel=Number of samples ($N$),
          ylabel=Lower bound ($\eta$),
          x label style={at={(axis description cs:0.4,-0.2)}},
          x tick label style={rotate=90,anchor=east},
          xmin=0,
          xmax=1000,
          ymin=0,
          ymax=1,
          every axis plot/.append style={line width=1pt},
          legend cell align={left},
          legend columns=1,
          legend pos=south east,
        ]
        
        \addplot[mark=none, color=MidnightBlue, mark size=1.3pt] table[x=N, y=b0.9, col sep=semicolon] {tikz/scenario-plots/eta_risk_fixedFraction=0.1.csv};
        
        \addplot[mark=none, color=BurntOrange, mark size=1.3pt] table[x=N, y=b0.99, col sep=semicolon] {tikz/scenario-plots/eta_risk_fixedFraction=0.1.csv};
        
        \addplot[mark=none, color=OliveGreen, mark size=1.3pt] table[x=N, y=b0.999, col sep=semicolon] {tikz/scenario-plots/eta_risk_fixedFraction=0.1.csv};
        
        \legend{{$\beta=0.9$},{$\beta=0.99$},{$\beta=0.999$}}
      \end{axis}
    \end{tikzpicture}

\caption{Fraction of violating samples is $\violating / N = 10\%$.}
\label{fig:scenario_fixedFraction=10}

\end{subfigure}
\hfill
\begin{subfigure}[b]{\linewidth}
    \begin{tikzpicture}
      \begin{axis}[
          width=\linewidth,
          height=4.5cm,
          ymajorgrids,
          grid style={dashed,gray!30},
          xlabel=Number of samples ($N$),
          ylabel=Lower bound ($\eta$),
          x label style={at={(axis description cs:0.4,-0.2)}},
          x tick label style={rotate=90,anchor=east},
          xmin=0,
          xmax=1000,
          ymin=0,
          ymax=1,
          every axis plot/.append style={line width=1pt},
          legend cell align={left},
          legend columns=1,
          legend pos=north east,
        ]
        
        \addplot[mark=none, color=MidnightBlue, mark size=1.3pt] table[x=N, y=b0.9, col sep=semicolon] {tikz/scenario-plots/eta_risk_fixedFraction=0.5.csv};
        
        \addplot[mark=none, color=BurntOrange, mark size=1.3pt] table[x=N, y=b0.99, col sep=semicolon] {tikz/scenario-plots/eta_risk_fixedFraction=0.5.csv};
        
        \addplot[mark=none, color=OliveGreen, mark size=1.3pt] table[x=N, y=b0.999, col sep=semicolon] {tikz/scenario-plots/eta_risk_fixedFraction=0.5.csv};
        
        \legend{{$\beta=0.9$},{$\beta=0.99$},{$\beta=0.999$}}
      \end{axis}
    \end{tikzpicture}

\caption{Fraction of violating samples is $\violating / N = 50\%$.}
\label{fig:scenario_fixedFraction=100}

\end{subfigure}

%% file: sections/6_scenario_reach.tex
\section{Derivation of the Main Results}
\label{sec:derivations}

In this section, we explain how we obtain  Theorems~\ref{theorem:reach_bound} and~\ref{theorem:reach_bound_relax}. 
Toward proving these theorems, we reformulate our problem statement into the domain of linear programs (LPs). 
First, we define the case where we account for all but a small fraction of the parameters instantiations $u \in \paramspace[\mdp]$ (recall that $\paramspace[\mdp]$ typically has infinite cardinality), which we formalize using a so-called \emph{chance-constrained LP}.
We remark that solving this chance-constrained LP directly is difficult~\cite{campi2018introduction}. 
Instead, we formalize our sampling-based approach, which is based on \emph{scenario optimization}, and which only considers a finite number of sampled parameters $u \in \mathcal{U}_N$.

\subsection{Chance-Constrained LP Reformulation}

Recall from \autoref{sec:Problem} that the problem is to compute a lower bound $\eta$ on the satisfaction probability $\satreachprob$. 
In other words, when sampling a parameter instantiation $u \in \paramspace[\mdp]$ according to probability measure $\PP$, compute a lower bound $\eta$ on the probability that $\mdp[u] \models \varphi$.
If the specification $\varphi$ has a comparison operator $\leq$ and a threshold $\lambda$ (e.g., $\varphi =\reachPropgT$), then the condition $\mdp[u] \models \varphi$ is equivalent to $\sol_{\pmdp}(u) \leq \lambda$.
As the solution function (\autoref{definition:valueFunction}) is a function of (only) the parameter instantiation, the solution $\sol_{\pmdp}(u)$ is also a random variable with probability measure $\PP$.
Thus, we can formalize the problem of finding a lower bound $\eta$ based on the following \emph{chance-constrained LP}:
\begin{subequations}
\begin{align}
	&\displaystyle \minimize_{\tau \geq 0}  \;\;	\tau 	\label{eq:chanceconstrained_obj}
	\\
	&\displaystyle \subjectto \,\,\, \Pr \Big\{ u \in \paramspace[\mdp] \,\, \Big\vert \,\, \sol_{\pmdp}(u) \leq \tau \Big\} \geq \eta.
	\label{eq:chanceconstrained_constraint}
\end{align}
\label{eq:chanceconstrained_mc}
\end{subequations}

\noindent
Then, the satisfaction probability $\satreachprob$ is lower bounded by $\eta$, given that the optimal solution $\tau^*$ to \eqref{eq:chanceconstrained_mc} is at most $\lambda$.

Similarly, if the specification $\varphi$ has a lower bound comparison operator $\geq$ and a threshold $\lambda$, we consider the following chance-constrained LP:
\begin{subequations}
\begin{align}
	&\displaystyle \maximize_{\tau \geq 0}  \;\;	\tau 	\label{eq:chanceconstrained_obj_max}
	\\
	&\displaystyle \subjectto \,\,\, \Pr \Big\{ u \in \paramspace[\mdp] \,\, \Big\vert \,\, \sol_{\pmdp}(u) \geq \tau \Big\} \geq \eta.
	\label{eq:chanceconstrained_constraint_max}
\end{align}
\label{eq:chanceconstrained_mc_max}
\end{subequations}

\noindent
Note that the differences between \eqref{eq:chanceconstrained_mc} and \eqref{eq:chanceconstrained_mc_max} are the optimization direction and the operator within the chance constraint.
Solving these chance-constrained problems is very hard in general, in particular because the probability distribution of the parameters is unknown~\cite{campi2018introduction}.

In what follows, we introduce our sampling-based approach to solve these problems with high confidence.
For brevity, we assume specifications with a lower bound comparison operator as in \eqref{eq:chanceconstrained_mc}, but modifying our results for other operators is straightforward.

\subsection{Scenario Optimization Program}
\label{subsec:scenario_based_verification}

Instead of solving the chance-constrained LP in \eqref{eq:chanceconstrained_mc} directly, we compute probably approximately correct lower bounds on the satisfaction probability based on \emph{scenario optimization}~\cite{DBLP:journals/tac/CalafioreC06,DBLP:journals/siamjo/CampiG08}.
Specifically, we replace the chance constraint~\eqref{eq:chanceconstrained_constraint}, which asks for the satisfaction of a certain fraction of a set of infinitely many constraints, with a finite number of \emph{hard constraints} that are induced by the sampled parameters $u \in \mathcal{U}_N$.
The resulting optimization problem is called a \emph{scenario program}~\cite{campi2018introduction} and is formulated as follows:
\begin{subequations}
\begin{align}
	&\displaystyle \minimize_{\tau \geq 0}  \;\;	\tau 	
	\label{eq:scenario_obj}
	\\
	&\displaystyle \subjectto \,\,\, \sol_{\pmdp}(u) \leq \tau \quad \forall u \in \mathcal{U}_N.
	\label{eq:scenario_constraint}
\end{align}
\label{eq:scenario_program}
\end{subequations}
\noindent
Upon solving scenario program \eqref{eq:scenario_program}, we obtain a unique optimal solution $\tau^* = \max_{u \in \mathcal{U}_N} \sol_{\pmdp}(u)$.
In \autoref{subsec:scenario:Basic}, we show that \autoref{theorem:reach_bound} follows from solving program \eqref{eq:scenario_program} directly, while \autoref{theorem:reach_bound_relax} corresponds to a setting where we deal with samples that violate the specification.

\subsection{Restriction to Satisfying Samples}
\label{subsec:scenario:Basic}

Consider the case where we directly solve the scenario program \eqref{eq:scenario_program}.
In this case, the following theorem, which is based on~\cite[Theorem 2.4]{DBLP:journals/siamjo/CampiG08}, enables us to compute a high-confidence lower bound on the satisfaction probability.

\begin{lemma}
    \label{lemma:reach_bound}
    Let uMDP $\vmdp$, a specification $\varphi$ with operator $\leq$, and the set $\mathcal{U}_N$ of $N \geq 1$ sampled parameters. 
    Let $\tau^*$ be the optimal solution of \eqref{eq:scenario_program}, and choose a confidence probability $\confidence \in (0,1)$.
    Then, it holds that 
    \begin{equation}
        \label{eq:lemma_reach_bound} 
        \probdist^N \Big\{
            \Pr \{ u \in \paramspace[\mdp] \,\, \vert \,\, \sol_{\pmdp}(u) \leq \tau^* \} \geq (1-\confidence)^{\frac{1}{N}} 
        \Big\} \geq \confidence.
    \end{equation}
\end{lemma}

\noindent
\autoref{lemma:reach_bound} states that with a probability of \emph{at least} $\confidence$, the probability that $\sol_{\pmdp}(u) \leq \tau^*$ for the next sampled parameter $u \in \paramspace[\mdp]$ is at least $(1-\confidence)^{\frac{1}{N}}$.
To derive \autoref{theorem:reach_bound} from \autoref{lemma:reach_bound}, we choose the (sample-dependent) specification threshold $\lambda(\mathcal{U}_N) \geq \tau^*$ after solving the scenario program.
Then, \autoref{theorem:reach_bound} follows directly by observing that $\satreachprob \geq \Pr \{ u \in \paramspace[\mdp] \, \vert \, \sol_{\pmdp}(u) \leq \tau^* \}$, since $\lambda(\mathcal{U}_N) \geq \tau^*$.
We provide the proof of \autoref{lemma:reach_bound}, and thus of \autoref{theorem:reach_bound}, in \autoref{sec:proofs}.

\subsection{Treatment of Violating Samples}
\label{subsec:scenario:ViolatingSamples}

We now derive \autoref{theorem:reach_bound_relax}, which assumes a fixed threshold $\lambda$.
In this case, we cannot guarantee a-priori that $\lambda \geq \tau^*$, because some samples may induce a reachability probability above $\lambda$, as in \autoref{fig:scenarioApproach_visual_right}.
Recall that $\violating$ is the number of samples that violate the specification.
Loosely speaking, we relax the constraints for these $\violating$ samples, and compute the maximum probability over the remaining samples $\mathcal{U}_{\satisfying} \subseteq \mathcal{U}_N$, which we write as $\tau^+ = \max_{u \in \mathcal{U}_{\satisfying}} \sol_{\pmdp}(u)$.
The following theorem is adapted from~\cite[Theorem 2.1]{campi2011sampling} and computes a high-confidence lower bound on the satisfaction probability, using the values of $\violating$ and $\tau^+$.

\begin{lemma}
    \label{lemma:reach_bound_relax}
    Let uMDP $\vmdp$, a specification $\varphi$ with operator $\leq$, and the set $\mathcal{U}_N$ of $N \geq 1$ sampled parameters.
    Fix a confidence probability $\confidence \in (0,1)$.
    Then, it holds~that
    \begin{equation}
        \probdist^N \Big\{ 
            \Pr \{ u \in \paramspace[\mdp] \vert \sol_{\pmdp}(u) \leq \tau^+ \} \geq t^*(\violating)
        \Big\} \geq \confidence,
        \label{eq:lemma_reach_bound_relax_2}
    \end{equation}
    where $t^*(N) = 0$ for $\violating = N$, and for $k = 0, \ldots, N-1$, $t^*(k)$ is the solution of
    \begin{equation}
        \frac{1 - \beta}{N} = \sum_{i=0}^{k} \binom Ni (1-t)^i t^{N-i}.
        \label{eq:lemma_reach_bound_relax_1}
    \end{equation} 
\end{lemma}

\noindent
\autoref{lemma:reach_bound_relax} asserts that with a probability of \emph{at least} $\confidence$, the probability that $\sol_{\pmdp}(u) \leq \tau^+$ for the next sampled parameter $u \in \paramspace[\mdp]$ is at least $t^*(\violating)$, given that $\violating$ samples violate the specification $\varphi$.
\autoref{theorem:reach_bound_relax} follows directly from \autoref{lemma:reach_bound_relax}, by observing that by construction, $\tau^+ \leq \lambda$.
We provide the proof of \autoref{lemma:reach_bound_relax}, and thus of \autoref{theorem:reach_bound_relax}, in \autoref{sec:proofs}.

We note that \eqref{eq:lemma_reach_bound_relax_1} is the cumulative distribution function of a beta distribution with $\violating + 1$ and $N - \violating$ degrees of freedom~\cite{campi2018introduction}, which can easily be solved numerically for $t$.
Moreover, we can speed up the computations at run-time, by tabulating the solutions to \eqref{eq:lemma_reach_bound_relax_1} for all relevant values of $N$, $\confidence$ and $\violating$ up front.

%% file: sections/7_experiments.tex
\section{Numerical Examples}\label{sec:numerical_examples}

We implemented our approach in Python using the model checker Storm~\cite{DBLP:conf/cav/DehnertJK017} to construct and analyze samples of MDPs.
Our implementation is available at \url{https://doi.org/10.5281/zenodo.6674059}
All computations ran on a computer with 32 3.7 GHz cores, and 64 GB of RAM.

First, we apply our method to the UAV motion planning example introduced in \autoref{sec:Example}.
Thereafter, we report on a set of well-known benchmarks used in parameter synthesis~\cite{DBLP:journals/corr/abs-1903-07993} that are, for instance, available on the website of the tools \tool{PARAM}~\cite{param_sttt} or part of the \tool{PRISM} benchmark suite~\cite{KNP12b}. 
We demonstrate that with our method, we can either specify a lower bound $\eta$ on the satisfaction probability and compute with what confidence probability $\confidence$ we can guarantee this lower bound, or we can do this in the opposite direction (i.e., specify $\eta$ and compute $\confidence$).

\subsection{UAV Motion Planning}

\paragraph{Setup.}
Recall the example from \autoref{sec:Example} of a UAV which needs to deliver a payload while avoiding obstacles.
The weather conditions are uncertain, and this uncertainty is reflected in the parameters of the uMDP.
For the distributions over parameter values, that is, over weather conditions, we consider the following three cases:

\begin{enumerate}
    \item we assume a uniform distribution over the different weather conditions in each zone;
    \item the probability for a weather condition inducing a wind direction that pushes the UAV northbound (i.e., into the positive $y$-direction) is twice as likely as in other directions;
    \item it is twice as likely to push the UAV westbound (i.e., into the negative $x$-direction).
\end{enumerate}

\paragraph{Trajectories.}
We depict example trajectories of the UAV for these three cases in Fig.~\ref{fig:drone}.
The trajectory depicted by the black line represents a simulated trajectory for the first case (uniform distribution), taking a direct route to reach the target area. 
Similarly, the trajectories shown by the dotted purple and dashed blue lines are simulated trajectories for the second (stronger northbound wind, i.e., positive $x$-direction) and third cases (stronger westbound wind, i.e., positive $y$-direction), respectively.
Under these two weather conditions, the UAV takes different paths toward the goal in order to account for the stronger wind.
In particular, for the case with northbound wind, we observe that the UAV is able to fly close to the obstacle at the right bottom.
By contrast, for the case with westbound wind, the UAV avoids to get close to this obstacle, as the wind may push the UAV into the obstacles, and then reaches the target area.

\input{results/table_results_UAV}

\paragraph{Bounds on satisfaction probabilities.}
We sample $N = 1\,000$ parameters for each case and consider different confidence probabilities $\confidence$ between $0.9$ and $0.9999$.
Specifically, we consider the specification $\varphi = \reachProp{0.9}{T}$, which is satisfied if the probability to safely reach the goal region is at least $90\%$.
For all three weather conditions, we compute the lower bounds $\eta$ on both the satisfaction probability $\satreachprob$ and the unsatisfaction probability $\falseprob$, using \autoref{theorem:reach_bound_relax}.

The results are presented in \autoref{table:UAV}.
The highest lower bound on the satisfaction probability is given by the first weather condition, and is $\eta = 0.911$ (for $\confidence=0.9$) and $\eta = 0.906$ (for $\confidence=0.9999$).
In other words, under a uniform distribution over the weather conditions, the UAV will (with a confidence of at least $\confidence = 0.9999$) satisfy the specification on at least $90.6\%$ of the days.
The second and third weather conditions lead to $\eta = 0.770$ and $\eta = 0.768$ (for $\confidence=0.9999$), respectively, showing that it is harder to navigate around the obstacles with non-uniform probability distributions over the parameters.
The average time to run a full iteration of our approach on this uMDP with $900$ parameters and around $10\,000$ states (i.e., performing the sampling, model checking, and computing the lower bounds $\eta$) with $5\,000$ parameter samples is $9.5$ minutes.

\input{results/table_model_info}

\subsection{Parameter Synthesis Benchmarks}
\paragraph{Setup.} 
In our second set of benchmarks, we adopt parametric MDPs and pMCs from~\cite{quatmann-et-al-atva-2016}. 
We adapt the \emph{Consensus} protocol~\cite{consensus} and the \emph{Bounded Retransmission Protocol} (brp)~\cite{HSV94,DBLP:conf/papm/DArgenioJJL01} to uMDPs; the \emph{Crowds Protocol} (crowds)~\cite{shmatikov2004probabilistic} and the \emph{NAND Multiplexing} benchmark (nand)~\cite{HJ02} become uMCs.
Essentially, the PLA technique from~\cite{quatmann-et-al-atva-2016} allows to approximate the percentage of instantiations that satisfy (or do not satisfy) a specification, while assuming a uniform distribution over the parameter space.
\autoref{tab:model_information} lists the specification checked $(\varphi)$ and the number of parameters, states, and transitions for all benchmarks.
Note that the satisfying regions reported in \autoref{tab:model_information} approximate $\satprob$, while the unsatisfying regions approximate $\falseprob$. 
We provide these numbers as a baseline only: the computation via PLA cannot scale to more then tens of parameters~\cite{quatmann-et-al-atva-2016} and cannot cope with unknown distributions.
For all benchmarks, we assume a uniform distribution over the parameters.

\begin{figure}[t]
    \subfloat[$N=1\,000$ samples.]{%
    \includegraphics[height=3.2cm]{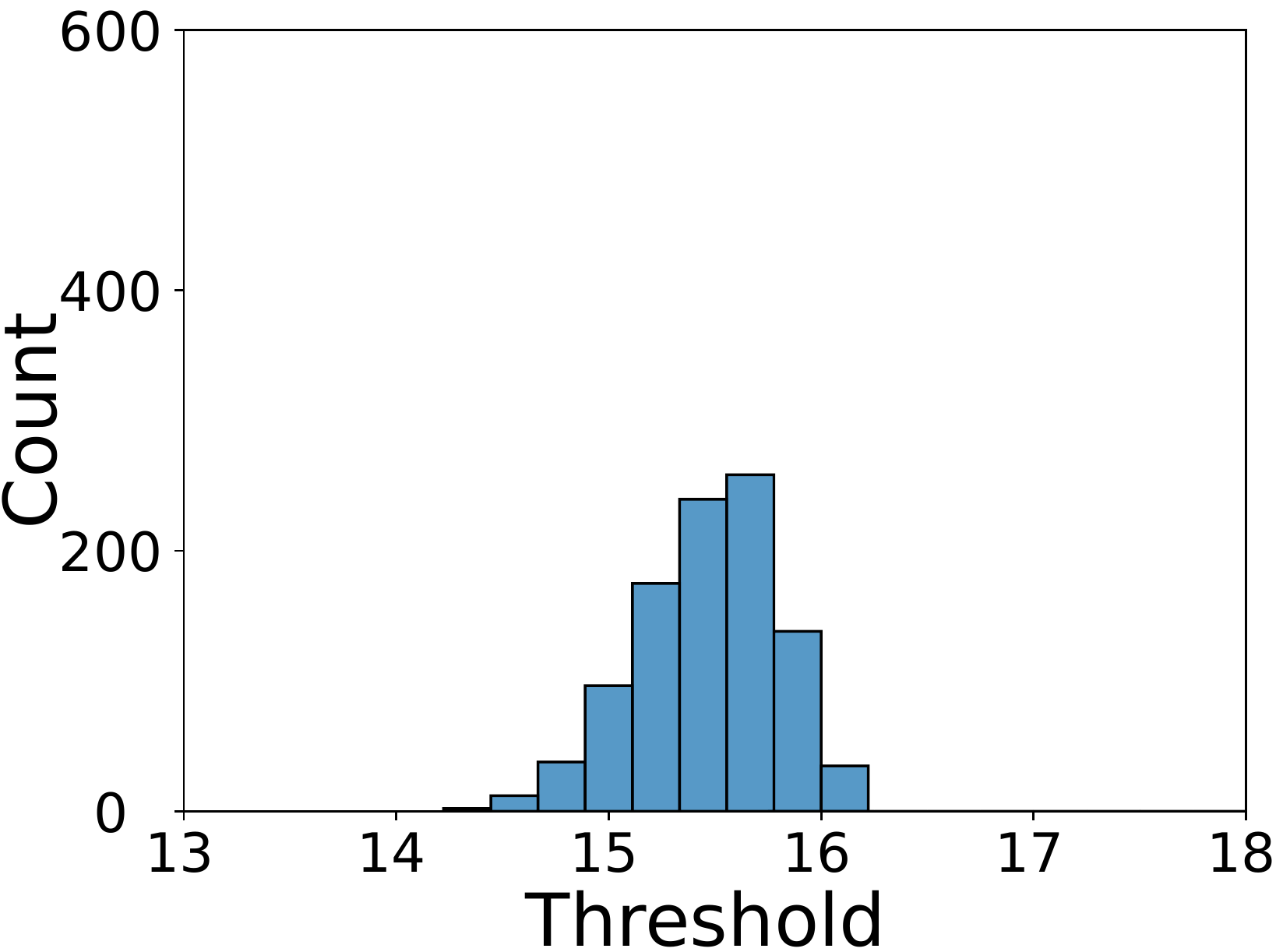}
    \label{fig:Histogram_a}}
    \subfloat[$N=10\,000$ samples.]{%
    \includegraphics[height=3.2cm]{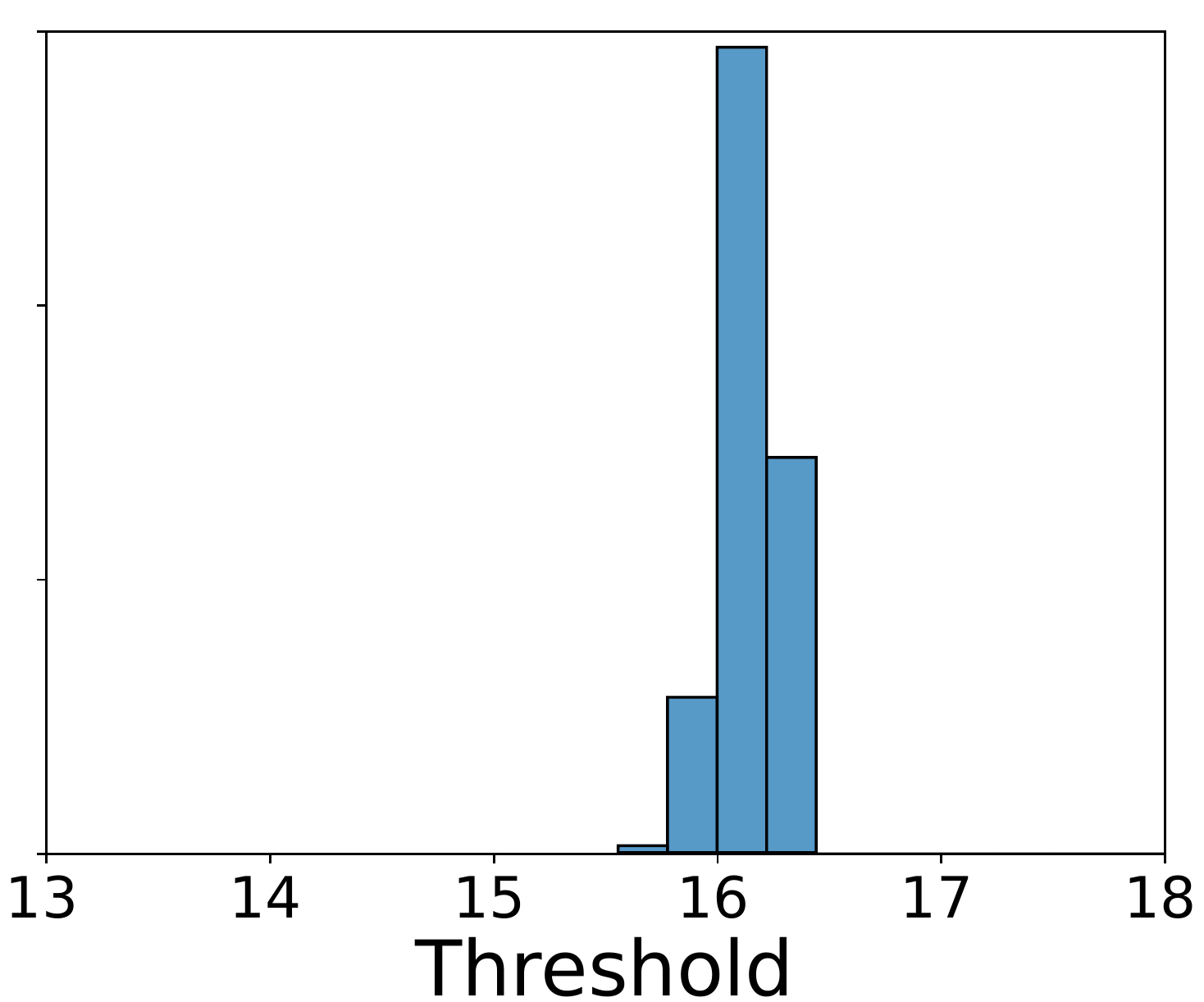}
    \label{fig:Histogram_b}}
    \caption{Histograms of the obtained thresholds $\lambda^*(\mathcal{U}_N)$.}
    \label{fig:Histogram}
\end{figure}

\paragraph{Specifications with variable thresholds $\lambda$.}
We demonstrate \autoref{theorem:reach_bound} on brp (16,5) with a variable threshold $\lambda^*(\mathcal{U}_N)$ in specification $\varphi = \mathbb{E}_{\leq \lambda^*(\mathcal{U})N)}(\finally T)$.
We use either $N=1\,000$ or $10\,000$ parameter samples and compute the tightest threshold $\lambda^*(\mathcal{U}_N)$ such that all samples are satisfying.
As explained in \autoref{ex:Samples}, this tightest threshold is $\lambda^*(\mathcal{U}_N) = \max_{u \in \mathcal{U}_N} \sol_{\pmdp}(u)$.
We repeat both cases ten thousand times and show a histogram of the obtained thresholds in \autoref{fig:Histogram}.
The corresponding lower bounds on the satisfaction probability (which only depend on $N$ and $\beta$) are $\eta=0.9954$ (for $N=1\,000$) and $\eta=0.9995$ (for $N=10\,000$).
We observe from \autoref{fig:Histogram} that for a higher number of samples, the thresholds are, on average, higher and their variability is lower.
These results are explainable, since the threshold is computed as the maximum of all solutions.

\input{results/table_results}

\paragraph{Computing $\confidence$ for a given $\eta$.}

We use \autoref{theorem:reach_bound_relax} to compute the confidence probabilities $\confidence$ that the approximate satisfying regions in \autoref{tab:model_information} are indeed a lower bound on the actual satisfaction probability $\satprob$ (and on $\falseprob$ for the unsatisfying regions).
Thus, we let $\eta$ be the approximate (un)satisfying regions in \autoref{tab:model_information}, and sample a desired number of parameters (between $N = 1\,000$ and $25\,000$).
For every instance, we then compute the confidence probability $\confidence$ using~\autoref{theorem:reach_bound_relax}.

For every benchmark and sample size, we report the average confidence probabilities $\confidence$ over $10$ full iterations of the same benchmark in~\autoref{table:table_confidence}.
Furthermore, we list the run time (in seconds) per $1\,000$ samples for each instance.
The results in Table~\ref{table:table_confidence} show that we can conclude with high confidence that the (un)satisfying regions are indeed a lower bound on the actual (un)satisfaction probabilities.
In particular for $N=25\,000$ samples, most confidence probabilities are very close to one. 
For example, for the crowds benchmark, instance (10,5) with $N=25\,000$, we obtain a confidence probability of $\confidence = 0.99943$ on the unsatisfying region of size $0.413$.
Thus, the probability that the approximate unsatisfying region of $0.413$ in \autoref{tab:model_information} is \emph{not} a lower bound on the actual unsatisfaction probability $\falseprob$ is less than $1-\confidence = 0.00057$.
Moreover, in line with \autoref{remark:independent}, larger models do not (in general) lead to worse confidence bounds (although model checking times do typically increase with the size of the MDP, cf. \autoref{table:table_confidence}).

The instance for which we obtained the worst confidence probability is the unsatisfying probability of nand (10,5), namely $\beta=0.975$.
Recomputing the results of~\cite{quatmann-et-al-atva-2016} with a much smaller tolerance revealed that the approximate unsatisfying region of $0.733$ was already a very tight lower bound (the best bound we were able to compute was $0.747$).
As such, we could only conclude with a confidence of $\beta=0.975$ that $\eta=0.733$ is a correct lower bound (as shown in \autoref{table:table_confidence}, for $N=25\,000$).

\input{results/table_results_fixBeta}

\paragraph{Computing $\eta$ for a given $\confidence$.}

Conversely, we can also use \autoref{theorem:reach_bound_relax} to compute the best lower bound $\eta$ on the (un)satisfaction probability that holds with at least a confidence probability $\confidence$.
For each benchmark, we sample $N=25\,000$ parameters and apply \autoref{theorem:reach_bound_relax} for increasing confidence probabilities $\confidence$.
We report the resulting bounds $\eta$ in~\autoref{table:table_satisfactionBound}.
We observe that the obtained values of $\eta$ are slightly more conservative (i.e., lower) for higher values of $\confidence$.
This observation is indeed intuitive: to reduce our chance $1-\confidence$ of obtaining an incorrect bound on the (un)satisfaction probability, the value of $\eta$ must be more conservative.
Moreover, increasing the confidence probability $\confidence$ only marginally reduces the obtained lower bound $\eta$.
For example, the obtained lower bound on the satisfaction probability for brp (256,5) with $\confidence = 0.9$ is $\eta = 0.07244$, while for $\confidence = 0.9999$, it is only reduced to $\eta = 0.07036$ (a reduction of only $0.21\%$). This observation confirms the important result of \autoref{coro:SampleComplexity}: a high confidence probability $\confidence$ can typically be obtained without sacrificing the tightness of the obtained lower bound $\eta$.

Recall that, based on \autoref{table:table_confidence}, we can only confirm the validity of the lower bound $\eta=0.733$ on the unsatisfying region for nand (10,5) with a confidence probability of $\beta=0.975$. 
Interestingly, \autoref{table:table_satisfactionBound} shows that we can guarantee a marginally weaker lower bound of $\eta = 0.73271$ with a confidence probability $\beta = 0.9999$.
In other words, by weakening our lower bound $\eta$ by an almost negligible amount of $0.0003$, we increase the confidence on the results from $97.51\%$ to a remarkable $99.99\%$.
This highlights that the confidence probability $\beta$ is extremely sensitive for the lower bound $\eta$, especially for high sample sizes $N$.

%% file: results/table_results_UAV.tex
{
\setlength{\tabcolsep}{4pt}
\begin{table*}[t]
\centering

\caption{Lower bounds $\eta$ on the (un)satisfaction probability for the UAV benchmark with $N=5\,000$ samples.}

\scalebox{1.0}{
\begin{tabular}{cccccccccccc}
	\hline
 \rule{0pt}{1.5ex}  
  Confidence probability
  & \multicolumn{2}{c}{{$\confidence = 0.9$}}
  & \multicolumn{2}{c}{{$\confidence = 0.99$}}
  & \multicolumn{2}{c}{{$\confidence = 0.999$}}
  & \multicolumn{2}{c}{{$\confidence = 0.9999$}}
   \\
 \cmidrule(lr){2-3} \cmidrule(lr){4-5} \cmidrule(lr){6-7} \cmidrule(lr){8-9}
  Weather condition
  & \multicolumn{1}{c}{$\eta$, sat} 
  & \multicolumn{1}{c}{$\eta$, unsat} 
  & \multicolumn{1}{c}{$\eta$, sat} 
  & \multicolumn{1}{c}{$\eta$, unsat} 
  & \multicolumn{1}{c}{$\eta$, sat} 
  & \multicolumn{1}{c}{$\eta$, unsat}  
  & \multicolumn{1}{c}{$\eta$, sat} 
  & \multicolumn{1}{c}{$\eta$, unsat}  
  \\
\hline
\hline
1. Uniform distribution & 0.91138 & 0.0583 & 0.90928 & 0.0567 & 0.90735 & 0.05528 & 0.90555 & 0.05398 \\
2. Stronger northbound wind & 0.77878 & 0.17483 & 0.77577 & 0.17217 & 0.77302 & 0.16978 & 0.77048 & 0.1676 \\
3. Stronger westbound wind & 0.7768 & 0.17664 & 0.77378 & 0.17397 & 0.77103 & 0.17157 & 0.76847 & 0.16938 \\
\hline
\end{tabular}
}

\label{table:UAV}

\end{table*}
}

%% file: results/table_model_info.tex
{
\setlength{\tabcolsep}{4pt}
\begin{table*}[t!]

\centering
\caption{Information for the benchmark instances and the approximate (un)satisfaction probabilities taken from~\cite{quatmann-et-al-atva-2016}.}

\scalebox{0.9}{
\begin{tabular}{cccccrrccrrr}
	\hline
 \rule{0pt}{1.5ex}  &  &  &  & 
  & \multicolumn{2}{c}{{Model size}}
  & \multicolumn{2}{c}{{Approximate (un)satisfying regions}}
   \\
  \cmidrule(lr){6-7} \cmidrule(lr){8-9} \cmidrule(lr){10-11}
  & benchmark
  & instance 
  & $\varphi$
  & \#pars
  & \multicolumn{1}{c}{\enskip\#states} 
  & \multicolumn{1}{c}{\#trans} 
  & \multicolumn{1}{c}{\enskip\enskip Satisfying region} 
  & \multicolumn{1}{c}{Unsatisfying region}  
  \\
  \hline
\hline
  \parbox[t]{3mm}{\multirow{9}{*}{\rotatebox[origin=c]{90}{\textbf{}}}}
  & \multirow{3}{*}{brp} \rule{0pt}{1.5ex}
  &     (256,5)     & $\reachPropg{0.5}{T}$ & 2 &     19\,720 &      26\,627 &  \enskip\enskip\enskip 0.055   &	0.898 \\
  & &     (16,5)      & $\mathbb{E}_{\leq 3}(\finally T)$ & 4 &      1\,304 &       1\,731 & \enskip\enskip\enskip 0.275   &   0.676     \\
  & &     (32,5)      & $\mathbb{E}_{\leq 3}(\finally T)$ & 4 &      2\,600 &       3\,459 & \enskip\enskip\enskip 0.232   &    0.718     \\
\cline{2-11}
  & \multirow{2}{*}{crowds} \rule{0pt}{2.5ex}
  &     (10,5)      & $\reachPropg{0.9}{T}$ & 2 &    104\,512 &     246\,082 & \enskip\enskip\enskip   0.537   &  0.413         \\
  & &     (15,7)      & $\reachPropg{0.9}{T}$ & 2 & 8\,364\,409 & 25\,108\,729 & \enskip\enskip\enskip 0.411   &  0.539  \\
\cline{2-11}
  & \multirow{2}{*}{nand}   \rule{0pt}{2.5ex}
  &     (10,5)      & $\reachProp{0.05}{T}$ & 2 &     35\,112 &       52\,647 &  \enskip\enskip\enskip 0.218   &  0.733      \\
  & &     (25,5)      & $\reachProp{0.05}{T}$ & 2 &    865\,592 &   1\,347\,047    & \enskip\enskip\enskip 0.206   &   0.744     \\ 
\cline{2-11}
  \parbox[t]{3mm}{\multirow{12}{*}{\rotatebox[origin=c]{90}{\textbf{}}}}
  & \multirow{2}{*}{consensus} \rule{0pt}{2.5ex}
  &      (2,2)      & $\reachProp{0.25}{T}$ & 2 &        272 &         492 & \enskip\enskip\enskip 0.280   &  0.669      \\
  & &      (4,2)      & $\reachProp{0.25}{T}$ & 4 &     22\,656 &      75\,232 &  \enskip\enskip\enskip 0.063   &  0.888     \\
\hline
\end{tabular}
}

\label{tab:model_information}
\end{table*}
}

%% file: results/table_results.tex
{
\setlength{\tabcolsep}{4pt}
\begin{table*}[t]
\centering

\caption{Average confidence probabilities $\beta$ for different sample sizes $N$, and run times per $1\,000$ samples. 
}
\scalebox{0.9}{
\begin{tabular}{cccccccccccccc}
	\hline
 \rule{0pt}{1.5ex}  & $\#$ samples  &  
  & \multicolumn{2}{c}{{1\,000}}
  & \multicolumn{2}{c}{{2\,500}} 
  & \multicolumn{2}{c}{{5\,000}}
  & \multicolumn{2}{c}{{10\,000}}
  & \multicolumn{2}{c}{{25\,000}}
  &
   \\
 \cmidrule(lr){4-5} \cmidrule(lr){6-7} \cmidrule(lr){8-9} \cmidrule(lr){10-11} \cmidrule(lr){12-13} 
  & benchmark
  & instance 
  & \multicolumn{1}{c}{$\beta$, sat} 
  & \multicolumn{1}{c}{$\beta$, unsat} 
  & \multicolumn{1}{c}{$\beta$, sat} 
  & \multicolumn{1}{c}{$\beta$, unsat} 
  & \multicolumn{1}{c}{$\beta$, sat} 
  & \multicolumn{1}{c}{$\beta$, unsat}
  & \multicolumn{1}{c}{$\beta$, sat} 
  & \multicolumn{1}{c}{$\beta$, unsat}
  & \multicolumn{1}{c}{$\beta$, sat} 
  & \multicolumn{1}{c}{$\beta$, unsat}
  & \multicolumn{1}{c}{Time (s)}  
  \\
\hline
\hline
  & \multirow{3}{*}{brp}  \rule{0pt}{3.0ex}
  &     (256,5)     & $0.42586$ & $0.14955$ & $0.91278$ & $0.53890$ & $0.99927$ & $0.91447$ & $1.00000$ & $0.99957$ & $1.00000$ & $1.00000$ & $1.296$ \\ 
  & &     (16,5)       & $0.05699$ & $0.03247$ & $0.22192$ & $0.10293$ & $0.62397$ & $0.35332$ & $0.95612$ & $0.92778$ & $1.00000$ & $1.00000$ & $0.341$ \\
  & &     (32,5)       & $0.05126$ & $0.07365$ & $0.21862$ & $0.27669$ & $0.50731$ & $0.64205$ & $0.89816$ & $0.91258$ & $1.00000$ & $1.00000$ & $0.344$ \\
\cline{2-14}
  & \multirow{2}{*}{crowds} 
  \rule{0pt}{3.0ex} 
  &     (10,5)        & $0.04770$ & $0.03339$ & $0.22113$ & $0.07921$ & $0.58451$ & $0.31034$ & $0.94009$ & $0.65727$ & $1.00000$ & $0.99943$ & $0.119$ \\
  & &     (15,7)      & $0.06568$ & $0.02839$ & $0.15451$ & $0.05446$ & $0.51860$ & $0.31260$ & $0.95223$ & $0.71819$ & $1.00000$ & $1.00000$ & $0.174$ \\
\cline{2-14}
  & \multirow{2}{*}{nand} 
    \rule{0pt}{3.0ex}   
  &     (10,5)      & $0.18263$ & $0.01101$ & $0.62057$ & $0.02775$ & $0.97510$ & $0.07370$ & $1.00000$ & $0.37567$ & $1.00000$ & $0.97509$ & $4.097$ \\
  & &     (25,5)    & $0.02938$ & $0.20327$ & $0.14312$ & $0.51272$ & $0.40151$ & $0.82369$ & $0.62267$ & $0.99994$ & $0.99884$ & $1.00000$ & $156.654$ \\ 
 \cline{2-14}
  & \multirow{2}{*}{consensus} 
    \rule{0pt}{3.0ex} 
  &      (2,2)      & $0.06282$ & $0.02833$ & $0.23683$ & $0.14101$ & $0.65357$ & $0.44217$ & $0.98990$ & $0.93097$ & $1.00000$ & $1.00000$ & $0.450$ \\
  & &      (4,2)    & $0.13668$ & $0.41820$ & $0.48546$ & $0.90556$ & $0.86663$ & $0.99999$ & $0.99998$ & $1.00000$ & $1.00000$ & $1.00000$ & $26.575$ \\ 
\hline
\end{tabular}
}

\label{table:table_confidence}

\end{table*}
}

%% file: results/table_results_fixBeta.tex
{
\setlength{\tabcolsep}{4pt}
\begin{table*}[t]
\centering

\caption{Lower bounds $\eta$ on the (un)satisfaction probability for $N = 25\,000$ samples.}

\scalebox{0.9}{
\begin{tabular}{ccccccccccccc}
	\hline
 \rule{0pt}{1.5ex}  & Confidence probability  &  
  & \multicolumn{2}{c}{{$\confidence = 0.9$}}
  & \multicolumn{2}{c}{{$\confidence = 0.99$}}
  & \multicolumn{2}{c}{{$\confidence = 0.999$}}
  & \multicolumn{2}{c}{{$\confidence = 0.9999$}}
   \\
 \cmidrule(lr){4-5} \cmidrule(lr){6-7} \cmidrule(lr){8-9} \cmidrule(lr){10-11} 
  & benchmark
  & instance 
  & \multicolumn{1}{c}{$\eta$, sat} 
  & \multicolumn{1}{c}{$\eta$, unsat} 
  & \multicolumn{1}{c}{$\eta$, sat} 
  & \multicolumn{1}{c}{$\eta$, unsat} 
  & \multicolumn{1}{c}{$\eta$, sat} 
  & \multicolumn{1}{c}{$\eta$, unsat}  
  & \multicolumn{1}{c}{$\eta$, sat} 
  & \multicolumn{1}{c}{$\eta$, unsat}  
  \\
\hline
\hline
  & \multirow{3}{*}{brp}  \rule{0pt}{3.0ex}
  &     (256,5)     &    $0.07244$ & $0.91221$ & $0.07168$ & $0.91135$ & $0.07099$ & $0.91056$ & $0.07036$ & $0.90982$ \\
  & &     (16,5)      &    $0.28787$ & $0.68619$ & $0.28653$ & $0.68481$ & $0.28531$ & $0.68353$ & $0.28417$ & $0.68234$ \\
  & &     (32,5)      &    $0.24356$ & $0.73176$ & $0.24229$ & $0.73044$ & $0.24113$ & $0.72922$ & $0.24005$ & $0.72808$ \\
\cline{2-11}
  & \multirow{2}{*}{crowds} 
  \rule{0pt}{3.0ex} 
  &     (10,5)      & $0.55106$ & $0.42091$ & $0.54957$ & $0.41945$ & $0.54821$ & $0.41810$ & $0.54695$ & $0.41685$ \\
& &     (15,7)      & $0.42397$ & $0.54798$ & $0.42250$ & $0.54650$ & $0.42115$ & $0.54514$ & $0.41990$ & $0.54387$ \\
\cline{2-11}
  & \multirow{2}{*}{nand} 
    \rule{0pt}{3.0ex}   
  &     (10,5)      & $0.23909$ & $0.73637$ & $0.23783$ & $0.73506$ & $0.23668$ & $0.73384$ & $0.23561$ & $0.73271$ \\
& &     (25,5)      & $0.20979$ & $0.76673$ & $0.20858$ & $0.76546$ & $0.20748$ & $0.76430$ & $0.20647$ & $0.76321$ \\
 \cline{2-11}
  & \multirow{2}{*}{consensus} 
    \rule{0pt}{3.0ex} 
  &      (2,2)      & $0.29383$ & $0.68009$ & $0.29248$ & $0.67870$ & $0.29125$ & $0.67742$ & $0.29010$ & $0.67622$ \\
  & &      (4,2)      & $0.07367$ & $0.91086$ & $0.07291$ & $0.91000$ & $0.07221$ & $0.90921$ & $0.07157$ & $0.90846$ \\

\hline
\end{tabular}
}

\label{table:table_satisfactionBound}

\end{table*}
}

%% file: sections/8_related.tex
\section{Discussion and Related work}
\label{sec:related}

The so-called~\emph{parameter synthesis} problem considers computing parameter values such that the induced non-parametric MDP satisfies the specification for some strategy.
Most works on parameter synthesis focus on finding one parameter value that satisfies the specification.
The approaches involve computing a rational function of the reachability probabilities~\cite{Daw04,param_sttt,DBLP:journals/corr/abs-1804-01872,DBLP:journals/iandc/BaierHHJKK20}, utilizing convex optimization~\cite{DBLP:conf/tacas/Cubuktepe0JKPPT17,atvaqcqp}, and sampling-based methods~\cite{chen2013model,meedeniya2014evaluating}.
The problem of whether there exists a value in the parameter space that satisfies a reachability specification is ETR-complete\footnote{The ETR satisfiability problem is to decide if there exists a satisfying assignment to the real variables in a Boolean combination of a set of polynomial inequalities. 
It is known that NP $\subseteq$ ETR $\subseteq$ PSPACE.}~\cite{winkler2019complexity}, and finding a satisfying parameter value is exponential in the number of parameters.

The work in~\cite{bacci2019model} considers the analysis of Markov models in the presence of uncertain rewards, utilizing statistical methods to reason about the probability of a parametric MDP satisfying an expected cost specification.
This approach is restricted to reward parameters and does not explicitly compute confidence bounds. The work in \cite{DBLP:conf/qest/PolgreenWHA16} obtains data-driven bounds on the parameter ranges and then uses parameter synthesis techniques to validate properties for all parameter values in this range.
Paper \cite{DBLP:conf/sigsoft/LlerenaBBSR18}~computes bounds on the long-run probability of satisfying a specification with probabilistic uncertainty for Markov chains.
Other related techniques include multi-objective model checking to maximize the average performance with probabilistic uncertainty sets~\cite{DBLP:conf/valuetools/Scheftelowitsch17}, sampling-based methods which minimize the \emph{regret} with uncertainty sets~\cite{DBLP:journals/jair/AhmedVLAJ17}, and Bayesian reasoning to compute parameter values that satisfy a metric temporal logic specification on a continuous-time Markov chain (CTMC)~\cite{DBLP:conf/tacas/BortolussiS18}.
Sampling-based methods similar to ours for verifying CTMCs with uncertain rates are developed in~\cite{Badings2022CAV}.
Finally, the work in \cite{arming2018parameter} considers a variant of the problem in this paper where parameter values cannot be observed and thus must be learned.
The paper formulates the strategy synthesis problem as an computationally harder partially observable Markov decision process (POMDP) synthesis problem and uses off-the-shelf point-based POMDP methods~\cite{pineau2003point,cassandra1997incremental}.

The works~\cite{puggelli2013polynomial,DBLP:conf/cdc/WolffTM12} consider the verification of MDPs with convex uncertainties. 
However, the uncertainty sets for different states in an MDP are restricted to be independent, which does not hold in our problem setting where we have parameter dependencies.

Uncertainties in MDPs have received quite some attention in the artificial intelligence and planning literature. 
Interval MDPs~\cite{puggelli2013polynomial,givan2000bounded} use probability intervals in the transition probabilities. Dynamic programming, robust value iteration and robust strategy iteration have been developed for MDPs with uncertain transition probabilities whose parameters are statistically independent, also referred to as rectangular, to find a strategy ensuring the highest expected total reward at a given confidence level~\cite{nilim2005robust,DBLP:conf/cdc/WolffTM12}. The work in~\cite{wiesemann2013robust} relaxes this independence assumption a bit and determines a strategy that satisfies a given performance with a pre-defined confidence provided an observation history of the MDP is given by using conic programming. State-of-the art exact methods can handle models of up to a few hundred of states~\cite{ho2018fast}. Multi-model MDPs~\cite{steimle2018multi} treat distributions over probability and cost parameters and aim at finding a single strategy maximizing a weighted value function. 
This problem is NP-hard for deterministic strategies and PSPACE-hard for history-dependent strategies.

%% file: sections/9_conclusion.tex
\section{Conclusion}
We have presented a new sampling-based approach to analyze uncertain Markov models. 
Theoretically, we have shown how to effectively and efficiently bound the probability that any randomly drawn sample satisfies a temporal logic specification. 
Furthermore, we have shown the computational tractability of our approaches by means of well-known benchmarks and a new, dedicated case study. 
In the future, we plan to exploit our approaches for more involved models such as Markov automata~\cite{DBLP:journals/eceasst/HatefiH12}. 
Another line of future work will be a closer integration with a parameter synthesis framework. 

%% file: sections/A_proofs.tex
\section{Proofs}
\label{sec:proofs}

In this section, we provide the proofs of our main theoretical contributions.
Since our theorems and lemmas are tailored to specifications $\varphi$ with comparison operator $\leq$, we also use this assumptions throughout the proofs.
The proofs for the case where we have a specification $\varphi$ with comparison operator $\geq$ are analogous to the difference between LPs~\eqref{eq:chanceconstrained_constraint} and~\eqref{eq:chanceconstrained_constraint_max}: we use $\maximize$ instead of $\minimize$, and the operator in the constraint changes sign.

\subsection{Proof of~\autoref{lemma:reach_bound} and~\autoref{theorem:reach_bound}}

We first prove~\autoref{lemma:reach_bound}, and then show that~\autoref{theorem:reach_bound} follows directly.
Let us rewrite LP~\eqref{eq:scenario_program} in a more compact way.
To this end, for a parameter sample $u \in \mathcal{U}_N$, let $C_{u}$ denote the interval of values for $\tau$, for which constraint~\eqref{eq:scenario_constraint} is satisfied.
Note that for a specification $\varphi$ with comparison operator $\leq$, we have $C_{u} = [\sol_{\pmdp}(u), \, +\infty)$, i.e., $C_u$ is lower bounded by the solution $\sol_{\pmdp}(u)$.
Using $C_{u}$, we reformulate the scenario program~\eqref{eq:scenario_program} as the equivalent program
\begin{equation}
\begin{split}
	\label{eq:scenario_program_compact}
	&\minimize_{\tau \geq 0}   \;\;	\tau
	\\
	&\subjectto \quad \tau \in \bigcap_{u \in \mathcal{U}_N} C_{u}.
\end{split}
\end{equation}
Note that~\eqref{eq:scenario_program_compact} is exactly in the form of the scenario program formulated in~\cite{DBLP:journals/siamjo/CampiG08}.
Let $\tau^*$ denote the optimal value to the scenario program with respect to sample set $\mathcal{U}_N$, and let $u$ be an independently sampled parameter from $\paramspace[\mdp]$ according to $\PP$.
Then, Theorem 2.4 of~\cite{DBLP:journals/siamjo/CampiG08} states that the \emph{cumulative distribution function} of the probability for the set $C_u$ associated with sampled parameter $u$ to \emph{violate} the optimal solution $\tau^*$, that is $\tau^* \notin C_u$, is written as follows:
\begin{equation}
\begin{split}
    \probdist^N \Big\{ \Pr\{  u \in \paramspace[\mdp] \, \vert \, & \tau^* \notin C_u \} > \violation \Big\} 
    \\
    & \leq 
    \sum_{i=0}^{d-1} 
    \begin{pmatrix}
		N \\
		i 
	\end{pmatrix}
	\violation^i (1-\violation)^{N-i},
\end{split}
\label{eq:reach_bound_proof1}
\end{equation}
where $\violation \in (0, 1)$ bounds the violation probability, and $d$ is the number of decision variables of~\eqref{eq:scenario_program_compact}.
Since we have $d=1$, and we are after the satisfaction probability (rather than the violation probability), we simplify~\eqref{eq:reach_bound_proof1} as
\begin{align}
    &\probdist^N \Big\{ \Pr\{ u \in \paramspace[\mdp] \, \vert \, \tau^* \in C_u \} < 1-\violation \Big\} 
    \leq 
    (1-\violation)^{N}
    \label{eq:reach_bound_proof2}
    \\
    &\probdist^N \Big\{ \Pr\{ u \in \paramspace[\mdp] \, \vert \, \tau^* \in C_u \} \geq 
    1-\violation \Big\} 
    \geq 
    1 - (1-\violation)^{N}.
    \label{eq:reach_bound_proof3}
\end{align}
Let $\confidence = 1 - (1-\violation)^N$, which implies that $\violation = 1-(1-\confidence)^{\frac{1}{N}}$.
Moreover, the event that $\tau^* \in C_u$ is equivalent to $\sol_{\pmdp}(u) \leq \tau^*$, so~\eqref{eq:reach_bound_proof3} reduces to
\begin{equation}
    \probdist^N \Big\{
            \Pr \{ u \in \paramspace[\mdp] \, \vert \, \sol_{\pmdp}(u) \leq \tau^* \} \geq (1-\confidence)^{\frac{1}{N}} 
        \Big\} \geq \confidence,
    \label{eq:reach_bound_proof4}
\end{equation}
which equals the desired expression in~\eqref{eq:lemma_reach_bound} for~\autoref{lemma:reach_bound}.

Finally, we show that~\autoref{theorem:reach_bound} follows from \autoref{lemma:reach_bound}.
Because $\tau^* \leq \lambda^*(\mathcal{U}_N)$, with $\lambda^*(\mathcal{U}_N)$ the sample-dependent threshold of specification $\varphi$, the inner probability in~\eqref{eq:reach_bound_proof4} is a lower bound on the satisfaction probability $\satreachprob$:
\begin{equation}
    \satreachprob \geq \Pr( u \in \paramspace[\mdp] \, \vert \, \tau^* \in C_u ). 
    \label{eq:reach_bound_proof5}
\end{equation}
By combining~\eqref{eq:reach_bound_proof4} with~\eqref{eq:reach_bound_proof5}, we find that
\begin{equation}
    \probdist^N \Big\{ \satreachprob \geq (1-\confidence)^{\frac{1}{N}} \Big\} \geq \confidence,
\end{equation}
and thus, the claim in~\autoref{theorem:reach_bound} follows.

\subsection{Proof of~\autoref{lemma:reach_bound_relax} and~\autoref{theorem:reach_bound_relax}}

We first prove~\autoref{lemma:reach_bound_relax}, and then show that~\autoref{theorem:reach_bound_relax} follows directly.
We modify the scenario program~\eqref{eq:scenario_program} as a scenario program with \emph{discarded samples}\cite{campi2011sampling}, which allows for the removal of undesirable constraints:
\begin{subequations}
\begin{align}
    &\displaystyle \minimize_{\tau \geq 0} \;\;	\tau	
	\label{eq:scenario_obj_relax}
	\\
	&\displaystyle \subjectto \,\,\, \sol_{\pmdp}(u) \leq \tau \quad \forall u \in \mathcal{U}_N \backslash \mathcal{Q},
	\label{eq:scenario_constraint_relax}
\end{align}
\label{eq:scenario_program_relax}
\end{subequations}
where we introduced the sample removal set $\mathcal{Q}$, which accounts for a subset of samples whose constraints have been discarded.
We explicitly write the dependency of the optimal solution $\tau^*_{\cardQ}$ on the number of discarded samples $\cardQ$.
Critically, samples are removed based on the following rule:
\begin{lemma}
    The sample removal set $\mathcal{Q} \subseteq \{ 1, \ldots, N \}$ is obtained by iteratively removing the \emph{active constraints} from~\eqref{eq:scenario_program_relax}, i.e. the samples $u \in \mathcal{U}_N$ for which $\sol_{\pmdp}(u) = \tau^*_{\cardQ}$.
    \label{lemma:Q}
\end{lemma}
Note that the active constraint may not be unique, e.g., if the solution $\sol_{\pmdp}(u_1) = \sol_{\pmdp}(u_2)$ for $u_1 \neq u_2$.
In that case, a suitable \emph{tie-break rule} may be used to select a constraint to discard, as discussed in~\cite{campi2008exact}.

The main difference between programs~\eqref{eq:scenario_program_relax} and~\eqref{eq:scenario_program} is that instead of enforcing the constraint for every sample $u \in \mathcal{U}_N$, we only enforce the constraint for a subset of samples $u \in \mathcal{U}_N \backslash \mathcal{Q}$.
Based on the solution to~\eqref{eq:scenario_program_relax}, Theorem 2.1 of~\cite{campi2011sampling} bounds the \emph{cumulative distribution function} of the violation probability, in a similar manner as the guarantees given by~\eqref{eq:reach_bound_proof1}:
\begin{equation}
\begin{split}
    \mathbb{P}^N \Big\{ & 
        \Pr\{ u \in \paramspace[\mdp] \, \vert \, \tau^*_{\cardQ} \notin C_u \} > \violation
    \Big\}
    \\
    & \leq 
    \begin{pmatrix}
		\mathcal{Q} + d - 1 \\
		\mathcal{Q}
	\end{pmatrix}
    \sum_{i=0}^{\cardQ + d - 1} \begin{pmatrix}
		N \\
		I
	\end{pmatrix} \epsilon^i (1-\epsilon)^{N-i}
    \\
    & = 
    \sum_{i=0}^{\cardQ} \begin{pmatrix}
		N \\
		I
	\end{pmatrix} \epsilon^i (1-\epsilon)^{N-i},
\end{split}
\label{eq:Proof_relax1}
\end{equation}
where $\epsilon \in (0,1)$ bounds the violation probability, $d = 1$ is the number of decision variables of~\eqref{eq:scenario_program_relax}, and $\cardQ$ is the cardinality of the sample removal set.
As our goal is to bound the satisfaction probability (rather than the violation probability), we define $t = 1-\epsilon$, and rewrite~\eqref{eq:Proof_relax1} as
\begin{align}
    \begin{split}
        \mathbb{P}^N \Big\{ 
            \Pr\{ u \in \paramspace[\mdp] \, \vert \, & \tau^*_{\cardQ} \in C_u \} < 1 - \violation
        \Big\}
        \\
        & \leq 
        \sum_{i=0}^{\cardQ} \begin{pmatrix}
    		N \\
    		I
    	\end{pmatrix} \epsilon^i (1-\epsilon)^{N-i}
    	\label{eq:Proof_relax2}
    \end{split}
    \\
    \begin{split}
    \mathbb{P}^N \Big\{ 
        \Pr\{ u \in \paramspace[\mdp] \, \vert \, & \tau^*_{\cardQ} \in C_u \} \geq t
    \Big\}
    \\
    & \geq 
    1 - \sum_{i=0}^{\cardQ} \begin{pmatrix}
		N \\
		I
	\end{pmatrix} (1-t)^i t^{N-i}.
	\label{eq:Proof_relax3}
    \end{split}
\end{align}
We equate the right-hand side of~\eqref{eq:Proof_relax3} to $1 - \frac{1 - \beta}{N}$, where $\beta \in (0,1)$ is a confidence probability (typically close to one):
\begin{equation}
\begin{split}
    \mathbb{P}^N \Big\{ 
        \Pr\{ u \in \paramspace[\mdp] \, \vert \, \tau^*_{\cardQ} \in C_u \} \geq t^*(\cardQ)
    \Big\} 
    \geq 
    1 - \frac{1 - \beta}{N},
	\label{eq:Proof_relax4}
\end{split}
\end{equation}
where $t^*(\cardQ)$ is the solution to
\begin{equation}
    1 - \frac{1 - \beta}{N} = 1 - \sum_{i=0}^{\cardQ} \begin{pmatrix}
		N \\
		I
	\end{pmatrix} (1-t)^i t^{N-i}.
	\label{eq:Proof_relax5}
\end{equation}
We divide the confidence level by $N$ to account for all $N$ possible values for $\cardQ$, ranging from $0$ to $N-1$.
The value of $\cardQ$ that is actually needed depends on the sample set at hand, and is, therefore, not known a-priori (i.e., before observing the actual samples).
For brevity, denote by $\mathcal{A}_n$ the event that 
\begin{equation}
    Pr\{ u \in \paramspace[\mdp] \, \vert \, \tau^*_n \in C_u \} \geq t^*(n).
\end{equation}
The probability for this event to hold is $\mathbb{P}^N \{ \mathcal{A}_n \} \geq 1 - \frac{1-\beta}{N}$, and its complement $\mathcal{A}_n'$ has a probability of $\mathbb{P}^N \{ \mathcal{A}_n' \} \leq \frac{1-\beta}{N}$.
Based on Boole's inequality, it holds that
\begin{equation}
    \mathbb{P}^N \Big\{ \bigcup_{i = 0}^{N-1} \mathcal{A}_n' \Big\} \leq 
    \sum_{i=0}^{N-1} \mathbb{P}^N \big\{ \mathcal{A}_n' \big\}
    \leq
    \frac{1-\beta}{N} N
    =
    1-\beta.
    \label{eq:UnionBound}
\end{equation}
Thus, the probability of the intersection of all events is
\begin{equation}
    \mathbb{P}^N \Big\{ \bigcap_{i = 0}^{N-1} \mathcal{A}_n \Big\} 
    = 
    1 - \mathbb{P}^N \Big\{ \bigcup_{i = 0}^{N-1} \mathcal{A}_n' \Big\}
    \geq \beta.
    \label{eq:UnionBound2}
\end{equation}
In other words, the bounds on the satisfaction probability given by \eqref{eq:Proof_relax4} hold \emph{simultaneously for all values} $\cardQ = 0, \ldots, N-1$ with a confidence probability of at least $\beta$.

After observing the samples at hand, we determine the actual value of $\violating$ and plug it as $\cardQ$ into~\eqref{eq:Proof_relax4}.
The probability that this expression holds cannot be smaller than that of the intersection of all events in~\eqref{eq:UnionBound2}.
Hence, we obtain
\begin{equation}
    \probdist^N \Big\{ 
        \Pr \{ u \in \paramspace[\mdp] \, \vert \, \sol_{\pmdp}(u) \leq \tau^*_{\cardQ} \} \geq t^*(\violating)
    \Big\} \geq \confidence.
    \label{eq:Proof_relax6}
\end{equation}
Recall from \autoref{subsec:scenario:ViolatingSamples} that $\tau^+ = \max_{u \in \mathcal{U}_{\satisfying}} \sol_{\pmdp}(u)$, which is, by construction, equivalent to $\tau^*_{\cardQ}$ under $\cardQ$ removed samples.
Thus,~\eqref{eq:Proof_relax6} is equivalent to~\eqref{eq:lemma_reach_bound_relax_2}, and the definition of $t^*(\cardQ)$ in~\eqref{eq:Proof_relax5} equals~\eqref{eq:lemma_reach_bound_relax_1}, so the claim in~\autoref{lemma:reach_bound_relax} follows.

Finally, to show that \autoref{theorem:reach_bound_relax} follows directly from \autoref{lemma:reach_bound_relax}, we note that $\tau^+ = \tau^*_{\cardQ} \leq \lambda$, so it must hold that
\begin{equation}
    \satreachprob \geq \Pr\{ u \in \paramspace[\mdp] \, \vert \, \sol_{\mdp}(u) \leq \tau^+ \}. 
    \label{eq:reach_bound_relax_proof2}
\end{equation}
By combining~\eqref{eq:Proof_relax6} with~\eqref{eq:reach_bound_relax_proof2}, we find that
\begin{equation}
    \Pr \Big\{ 
            \satreachprob \, \geq \, t^*(\violating)
    \Big\} \geq \confidence,
\end{equation}
and thus, we conclude the proof of~\autoref{theorem:reach_bound_relax}.

%% file: main_STTT.bbl
\begin{thebibliography}{10}
\providecommand{\url}[1]{{#1}}
\providecommand{\urlprefix}{URL }
\expandafter\ifx\csname urlstyle\endcsname\relax
  \providecommand{\doi}[1]{DOI~\discretionary{}{}{}#1}\else
  \providecommand{\doi}{DOI~\discretionary{}{}{}\begingroup
  \urlstyle{rm}\Url}\fi

\bibitem{DBLP:journals/jair/AhmedVLAJ17}
Ahmed, A., Varakantham, P., Lowalekar, M., Adulyasak, Y., Jaillet, P.:
  {Sampling Based Approaches for Minimizing Regret in Uncertain Markov Decision
  Processes (MDPs)}.
\newblock J. Artif. Intell. Res. \textbf{59}, 229--264 (2017)

\bibitem{arming2018parameter}
Arming, S., Bartocci, E., Chatterjee, K., Katoen, J.P., Sokolova, A.:
  {Parameter-Independent Strategies for pMDPs via POMDPs}.
\newblock In: {QEST}, pp. 53--70. Springer (2018)

\bibitem{consensus}
Aspnes, J., Herlihy, M.: {Fast Randomized Consensus Using Shared Memory}.
\newblock Journal of Algorithms \textbf{15}(1), 441--460 (1990)

\bibitem{bacci2019model}
Bacci, G., Hansen, M., Larsen, K.G.: {Model Checking Constrained Markov Reward
  Models with Uncertainties}.
\newblock In: {QEST}, pp. 37--51 (2019)

\bibitem{Badings2022CAV}
Badings, T.S., Jansen, N., Junges, S., Stoelinga, M., Volk, M.: Sampling-based
  verification of ctmcs with uncertain rates.
\newblock In: International Conference on Computer Aided Verification (to
  appear). Springer (2022)

\bibitem{DBLP:journals/iandc/BaierHHJKK20}
Baier, C., Hensel, C., Hutschenreiter, L., Junges, S., Katoen, J.P., Klein, J.:
  Parametric {M}arkov chains: {PCTL} complexity and fraction-free {G}aussian
  elimination.
\newblock Inf. Comput. \textbf{272}, 104504 (2020)

\bibitem{BK08}
Baier, C., Katoen, J.P.: Principles of Model Checking.
\newblock MIT Press (2008)

\bibitem{Book_Basu_RAAlgorithms}
Basu, S., Pollack, R., Roy, M.: {Algorithms in Real Algebraic Geometry}.
\newblock Springer (2010)

\bibitem{bertsekas2000introduction}
Bertsekas, D.P., Tsitsiklis, J.N.: Introduction to probability.
\newblock Athena Scientinis (2000)

\bibitem{DBLP:conf/tacas/BortolussiS18}
Bortolussi, L., Silvetti, S.: Bayesian statistical parameter synthesis for
  linear temporal properties of stochastic models.
\newblock In: {TACAS} {(2)}, \emph{Lecture Notes in Computer Science}, vol.
  10806, pp. 396--413. Springer (2018)

\bibitem{boyd_convex_optimization}
Boyd, S., Vandenberghe, L.: Convex Optimization.
\newblock Cambridge University Press, New York, NY, USA (2004)

\bibitem{DBLP:journals/tac/CalafioreC06}
Calafiore, G.C., Campi, M.C.: {The Scenario Approach to Robust Control Design}.
\newblock {IEEE} Trans. Automat. Contr. \textbf{51}(5), 742--753 (2006)

\bibitem{DBLP:journals/siamjo/CampiG08}
Campi, M.C., Garatti, S.: {The Exact Feasibility of Randomized Solutions of
  Uncertain Convex Programs}.
\newblock {SIAM} Journal on Optimization \textbf{19}(3), 1211--1230 (2008)

\bibitem{campi2008exact}
Campi, M.C., Garatti, S.: {The Exact Feasibility of Randomized Solutions of
  Uncertain Convex Programs}.
\newblock SIAM Journal on Optimization \textbf{19}(3), 1211--1230 (2008)

\bibitem{campi2011sampling}
Campi, M.C., Garatti, S.: {A Sampling-and-Discarding Approach to
  Chance-Constrained Optimization: Feasibility and Optimality}.
\newblock Journal of Optimization Theory and Applications \textbf{148}(2),
  257--280 (2011)

\bibitem{campi2018introduction}
Campi, M.C., Garatti, S.: Introduction to the Scenario Approach.
\newblock SIAM (2018)

\bibitem{cassandra1997incremental}
Cassandra, A., Littman, M.L., Zhang, N.L.: {Incremental Pruning: A Simple,
  Fast, Exact Method for Partially Observable Markov Decision Processes}.
\newblock In: {UAI}, pp. 54--61 (1997)

\bibitem{chen2013model}
Chen, T., Hahn, E.M., Han, T., Kwiatkowska, M., Qu, H., Zhang, L.: {Model
  Repair for {M}arkov Decision Processes}.
\newblock In: TASE, pp. 85--92. IEEE CS (2013)

\bibitem{DBLP:conf/tacas/Cubuktepe0JKT20}
Cubuktepe, M., Jansen, N., Junges, S., Katoen, J., Topcu, U.: Scenario-based
  verification of uncertain mdps.
\newblock In: {TACAS} {(1)}, \emph{Lecture Notes in Computer Science}, vol.
  12078, pp. 287--305. Springer (2020)

\bibitem{DBLP:conf/tacas/Cubuktepe0JKPPT17}
Cubuktepe, M., Jansen, N., Junges, S., Katoen, J.P., Papusha, I., Poonawala,
  H.A., Topcu, U.: {Sequential Convex Programming for the Efficient
  Verification of Parametric MDPs}.
\newblock In: {TACAS} {(2)}, \emph{LNCS}, vol. 10206, pp. 133--150 (2017)

\bibitem{atvaqcqp}
Cubuktepe, M., Jansen, N., Junges, S., Katoen, J.P., Topcu, U.: Synthesis in
  {pMDPs}: {A} tale of 1001 parameters.
\newblock In: {ATVA}, \emph{{LNCS}}, vol. 11138, pp. 160--176. Springer (2018)

\bibitem{DBLP:conf/papm/DArgenioJJL01}
D'Argenio, P.R., Jeannet, B., Jensen, H.E., Larsen, K.G.: Reachability analysis
  of probabilistic systems by successive refinements.
\newblock In: {PAPM-PROBMIV}, \emph{LNCS}, vol. 2165, pp. 39--56. Springer
  (2001)

\bibitem{Daw04}
Daws, C.: {Symbolic and Parametric Model Checking of Discrete-Time {{M}arkov}}
  chains.
\newblock In: ICTAC, \emph{LNCS}, vol. 3407, pp. 280--294. Springer (2004)

\bibitem{dehnert-et-al-cav-2015}
Dehnert, C., Junges, S., Jansen, N., Corzilius, F., Volk, M., Bruintjes, H.,
  Katoen, J., {\'{A}}brah{\'{a}}m, E.: {PROPhESY: {A} PRObabilistic ParamEter
  SYnthesis Tool}.
\newblock In: {CAV} {(1)}, \emph{LNCS}, vol. 9206, pp. 214--231. Springer
  (2015)

\bibitem{DBLP:conf/cav/DehnertJK017}
Dehnert, C., Junges, S., Katoen, J., Volk, M.: {A Storm is Coming: {A} Modern
  Probabilistic Model Checker}.
\newblock In: {CAV} {(2)}, \emph{LNCS}, vol. 10427, pp. 592--600. Springer
  (2017)

\bibitem{DBLP:conf/lata/DelahayeLLPW11}
Delahaye, B., Larsen, K.G., Legay, A., Pedersen, M.L., Wasowski, A.: Decision
  problems for interval {M}arkov chains.
\newblock In: {LATA}, \emph{{LNCS}}, vol. 6638, pp. 274--285. Springer (2011)

\bibitem{DBLP:journals/corr/abs-1804-01872}
Gainer, P., Hahn, E.M., Schewe, S.: Incremental verification of parametric and
  reconfigurable {M}arkov chains.
\newblock In: {QEST}, \emph{LNCS}, vol. 11024, pp. 140--156. Springer (2018)

\bibitem{GarattiCampi2021}
Garatti, S., Campi, M.C.: The risk of making decisions from data through the
  lens of the scenario approach.
\newblock IFAC-PapersOnLine \textbf{54}(7), 607--612 (2021)

\bibitem{givan2000bounded}
Givan, R., Leach, S., Dean, T.: {Bounded-Parameter Markov Decision Processes}.
\newblock Artificial Intelligence \textbf{122}(1-2), 71--109 (2000)

\bibitem{param_sttt}
Hahn, E.M., Hermanns, H., Zhang, L.: {Probabilistic Reachability for Parametric
  Markov Models}.
\newblock STTT \textbf{13}(1), 3--19 (2010)

\bibitem{HJ02}
Han, J., Jonker, P.: {A System Architecture Solution for Unreliable
  Nanoelectronic Devices}.
\newblock IEEE Transactions on Nanotechnology \textbf{1}, 201--208 (2002)

\bibitem{DBLP:journals/eceasst/HatefiH12}
Hatefi, H., Hermanns, H.: {Model Checking Algorithms for Markov Automata}.
\newblock {ECEASST} \textbf{53} (2012)

\bibitem{DBLP:conf/aaai/Haussler90}
Haussler, D.: Probably approximately correct learning.
\newblock In: {AAAI}, pp. 1101--1108. {AAAI} Press / The {MIT} Press (1990)

\bibitem{HSV94}
Helmink, L., Sellink, M., Vaandrager, F.: {Proof-Checking a Data Link
  Protocol}.
\newblock In: TYPES, \emph{LNCS}, vol. 806, pp. 127--165. Springer (1994)

\bibitem{ho2018fast}
Ho, C.P., Petrik, M.: {Fast Bellman Updates for Robust MDPs}.
\newblock In: ICML (2018)

\bibitem{DBLP:journals/corr/abs-1903-07993}
Junges, S., {\'{A}}brah{\'{a}}m, E., Hensel, C., Jansen, N., Katoen, J.,
  Quatmann, T., Volk, M.: {Parameter Synthesis for Markov Models}.
\newblock CoRR \textbf{abs/1903.07993} (2019)

\bibitem{KNP11}
Kwiatkowska, M., Norman, G., Parker, D.: {\tool{PRISM} 4.0: Verification of
  Probabilistic Real-Time Systems}.
\newblock In: CAV, \emph{LNCS}, vol. 6806, pp. 585--591. Springer (2011)

\bibitem{KNP12b}
Kwiatkowska, M., Norman, G., Parker, D.: {The \tool{PRISM} Benchmark Suite}.
\newblock In: QEST, pp. 203--204. IEEE CS (2012)

\bibitem{DBLP:conf/sigsoft/LlerenaBBSR18}
Llerena, Y.R.S., B{\"{o}}hme, M., Br{\"{u}}nink, M., Su, G., Rosenblum, D.S.:
  {Verifying the Long-run Behavior of Probabilistic System Models in the
  Presence of Uncertainty}.
\newblock In: {ESEC/SIGSOFT} {FSE}, pp. 587--597. {ACM} (2018)

\bibitem{mcallister2012motion}
McAllister, R., Peynot, T., Fitch, R., Sukkarieh, S.: {Motion Planning and
  Stochastic Control with Experimental Validation on a Planetary Rover}.
\newblock In: {IROS}, pp. 4716--4723. IEEE (2012)

\bibitem{meedeniya2014evaluating}
Meedeniya, I., Moser, I., Aleti, A., Grunske, L.: {Evaluating Probabilistic
  Models with Uncertain Model Parameters}.
\newblock Software \& Systems Modeling \textbf{13}(4), 1395--1415 (2014)

\bibitem{nilim2005robust}
Nilim, A., El~Ghaoui, L.: {Robust Control of Markov Decision Processes with
  Uncertain Transition Matrices}.
\newblock Operations Research \textbf{53}(5), 780--798 (2005)

\bibitem{Papaefthymiou2008}
Papaefthymiou, G., Kl{\"{o}}ckl, B.: {MCMC for wind power simulation}.
\newblock IEEE Transactions on Energy Conversion \textbf{23}(1), 234--240
  (2008)

\bibitem{pineau2003point}
Pineau, J., Gordon, G., Thrun, S.: {Point-Based Value Iteration: an Anytime
  Algorithm for POMDPs}.
\newblock In: {IJCAI}, pp. 1025--1030 (2003)

\bibitem{Pnu77}
Pnueli, A.: {The Temporal Logic of Programs}.
\newblock In: FOCS, pp. 46--57 (1977)

\bibitem{DBLP:conf/qest/PolgreenWHA16}
Polgreen, E., Wijesuriya, V.B., Haesaert, S., Abate, A.: Data-efficient
  bayesian verification of parametric markov chains.
\newblock In: {QEST}, \emph{Lecture Notes in Computer Science}, vol. 9826, pp.
  35--51. Springer (2016)

\bibitem{puggelli2013polynomial}
Puggelli, A., Li, W., Sangiovanni-Vincentelli, A.L., Seshia, S.A.:
  {Polynomial-Time Verification of PCTL Properties of MDPs with Convex
  Uncertainties}.
\newblock In: CAV, pp. 527--542. Springer (2013)

\bibitem{puterman2014markov}
Puterman, M.L.: {Markov Decision Processes: Discrete Stochastic Dynamic
  Programming}.
\newblock John Wiley \& Sons (2014)

\bibitem{quatmann-et-al-atva-2016}
Quatmann, T., Dehnert, C., Jansen, N., Junges, S., Katoen, J.P.: {Parameter
  Synthesis for Markov Models: Faster Than Ever}.
\newblock In: ATVA, \emph{LNCS}, vol. 9938, pp. 50--67 (2016)

\bibitem{russell2016artificial}
Russell, S.J., Norvig, P.: Artificial Intelligence: {A} Modern Approach (4th
  Edition).
\newblock Pearson (2020)

\bibitem{DBLP:conf/valuetools/Scheftelowitsch17}
Scheftelowitsch, D., Buchholz, P., Hashemi, V., Hermanns, H.: {Multi-Objective
  Approaches to Markov Decision Processes with Uncertain Transition
  Parameters}.
\newblock In: {VALUETOOLS}, pp. 44--51 (2017)

\bibitem{shmatikov2004probabilistic}
Shmatikov, V.: {Probabilistic Analysis of an Anonymity System}.
\newblock Journal of Computer Security \textbf{12}(3-4), 355--377 (2004)

\bibitem{steimle2018multi}
Steimle, L.N., Kaufman, D.L., Denton, B.T.: {Multi-Model Markov Decision
  Processes}.
\newblock Optimization Online  (2018)

\bibitem{sutton2018reinforcement}
Sutton, R.S., Barto, A.G.: {Reinforcement Learning: An Introduction}.
\newblock MIT Press (2018)

\bibitem{wiesemann2013robust}
Wiesemann, W., Kuhn, D., Rustem, B.: {Robust Markov Decision Processes}.
\newblock Mathematics of Operations Research \textbf{38}(1), 153--183 (2013)

\bibitem{winkler2019complexity}
Winkler, T., Junges, S., P{\'{e}}rez, G.A., Katoen, J.P.: On the complexity of
  reachability in parametric {M}arkov decision processes.
\newblock In: {CONCUR}, \emph{LIPIcs}, vol. 140, pp. 14:1--14:17. Schloss
  Dagstuhl - Leibniz-Zentrum f{\"{u}}r Informatik (2019)

\bibitem{DBLP:conf/cdc/WolffTM12}
Wolff, E.M., Topcu, U., Murray, R.M.: Robust control of uncertain {M}arkov
  decision processes with temporal logic specifications.
\newblock In: {CDC}, pp. 3372--3379. {IEEE} (2012)

\end{thebibliography}
